\documentclass[12pt,preprint]{aastex}

\shorttitle{NGC 6342 and NGC 6366}
\shortauthors{Johnson et al.}

\newcommand\iso[2]{$^{\rm #1}$#2}

\begin{document}

\title{THE CHEMICAL COMPOSITION OF RED GIANT BRANCH STARS IN THE GALACTIC 
GLOBULAR CLUSTERS NGC 6342 AND NGC 6366}

\author{
Christian I. Johnson\altaffilmark{1,2},
Nelson Caldwell\altaffilmark{1},
R. Michael Rich\altaffilmark{3}, 
Catherine A. Pilachowski\altaffilmark{4}, and
Tiffany Hsyu\altaffilmark{5}
}

\altaffiltext{1}{Harvard--Smithsonian Center for Astrophysics, 60 Garden
Street, MS--15, Cambridge, MA 02138, USA; cjohnson@cfa.harvard.edu; 
ncaldwell@cfa.harvard.edu}

\altaffiltext{2}{Clay Fellow}

\altaffiltext{3}{Department of Physics and Astronomy, UCLA, 430 Portola Plaza,
Box 951547, Los Angeles, CA 90095-1547, USA; rmr@astro.ucla.edu}

\altaffiltext{4}{Astronomy Department, Indiana University Bloomington, Swain
West 319, 727 East 3rd Street, Bloomington, IN 47405--7105, USA;
cpilacho@indiana.edu}

\altaffiltext{5}{UCO/Lick Observatory, University of California, 1156 High 
Street, Santa Cruz, CA 95064, USA; thsyu@ucsc.edu}

\begin{abstract}

We present radial velocities and chemical abundances for red giant branch stars 
in the Galactic bulge globular clusters NGC 6342 and NGC 6366.  The velocities
and abundances are based on measurements of high resolution (R $\ga$ 20,000) 
spectra obtained with the MMT--Hectochelle and WIYN--Hydra spectrographs.  We 
find that NGC 6342 has a heliocentric radial velocity of 
$+$112.5 km s$^{\rm -1}$ ($\sigma$ = 8.6 km s$^{\rm -1}$), NGC 6366 has a 
heliocentric radial velocity of --122.3 km s$^{\rm -1}$ ($\sigma$ = 1.5 km 
s$^{\rm -1}$), and that both clusters have nearly identical metallicities 
([Fe/H] $\approx$ --0.55).  NGC 6366 shows evidence of a moderately extended 
O--Na anti--correlation, but more data are needed for NGC 6342 to determine if 
this cluster also exhibits the typical O--Na relation likely found in all 
other Galactic globular clusters.  The two clusters are distinguished from 
similar metallicity field stars as having larger [Na/Fe] spreads and enhanced 
[La/Fe] ratios, but we find that NGC 6342 and NGC 6366 display $\alpha$ and
Fe--peak element abundance patterns that are typical of other metal--rich
([Fe/H] $>$ --1) inner Galaxy clusters.  However, the median [La/Fe] abundance
may vary from cluster--to--cluster.

\end{abstract}


\section{INTRODUCTION}

Globular cluster systems offer insights into a galaxy's chemical
evolution, star formation history, dynamical evolution, and merger history.
In the Milky Way, globular clusters are often categorized based on 
characteristics such as chemical composition, age, horizontal branch 
morphology, and kinematics.  The observed metallicity distribution function of 
Galactic globular clusters is largely bimodal with approximately 2/3 of 
clusters belonging to a metal--poor group that peaks near 
[Fe/H]\footnote{[A/B]$\equiv$log(N$_{\rm A}$/N$_{\rm B}$)$_{\rm star}$--log(N$_{\rm A}$/N$_{\rm B}$)$_{\sun}$ and log $\epsilon$(A)$\equiv$log(N$_{\rm A}$/N$_{\rm H}$)+12.0 for elements A and B.} $\sim$ --1.5 and 1/3 of clusters belonging 
to a metal--rich group that peaks near [Fe/H] $\sim$ --0.5 (e.g., Freeman \& 
Norris 1981; Zinn 1985; Bica et al. 2006).  Furthermore, while metal--poor 
globular clusters are distributed across a wide range of galactocentric radii 
and are mostly associated with the Galactic halo, metal--rich clusters form a 
more flattened population that is concentrated near the 
inner few kpc of the Galaxy (e.g., Zinn 1985; van den Bergh 2003; Rossi et al. 
2015).  Recent work suggests that a majority of the inner Galaxy globular 
clusters with [Fe/H] $\ga$ --1 are members of the Galactic bulge/bar system 
(e.g., Minniti 1995; C{\^o}t{\'e} 1999; Rossi et al. 2015).  Interestingly, 
some age--metallicity relations find that the central metal--rich globular 
cluster population may even be coeval with, and in some cases older than 
(e.g., NGC 6522; Barbuy et al. 2009), some of the more metal--poor halo 
clusters (e.g., Mar{\'{\i}}n--Franch et al. 2009; Forbes \& Bridges 2010).
In contrast, VandenBerg et al. (2013) find that clusters with [Fe/H] $>$ --1
are younger than those with [Fe/H] $<$ --1, and also do not find a strong
correlation between galactocentric distance and age.

Although inner Galaxy globular clusters are not as extensively studied as their
halo counterparts (e.g., see reviews by Kraft 1994; Gratton et al. 2004; 
Gratton et al. 2012), several bulge clusters are known to exhibit unusual 
chemical and/or morphological characteristics.  For example, Haute--Provence 1 
(HP--1) is located near the Galactic center and is relatively metal--rich at
[Fe/H] = --1, but the cluster contains a prominent blue horizontal branch and 
no red horizontal branch stars (Ortolani et al. 1997; Barbuy et al. 2006; 
Ortolani et al. 2011).  Similarly, the bulge clusters NGC 6388 and NGC 6441 
have [Fe/H] $\sim$ --0.4 (e.g., Gratton et al. 2006; Carretta et al. 2007; 
Gratton et al. 2007; Origlia et al. 2008), anomalous red giant branch (RGB) 
bumps (Nataf et al. 2013), and dominant red horizontal branches accompanied by 
very extended blue horizontal branches (e.g., Rich et al. 1997; Bellini et al. 
2013).  For HP--1, NGC 6388, and NGC 6441, the presence of a significant 
population of blue horizontal branch stars is not expected given the clusters' 
metallicities.  In a similar sense, the bulge globular clusters NGC 6440 and 
NGC 6569 exhibit double red clumps that are so far observed only in near 
infrared color--magnitude diagrams (Mauro et al. 2012).  The underlying cause
of the double red clump feature in these clusters is not yet clear, but in 
Terzan 5 a double red clump has been linked to multiple stellar populations 
with metallicities that range from [Fe/H] $\sim$ --0.8 to $+$0.3 (Ferraro et 
al. 2009; Origlia et al. 2011; Origlia et al. 2013).  Detailed spectroscopic
analyses by Origlia et al. (2011; 2013) have further revealed that the 
chemical composition of each population appears to match that found in bulge 
field stars, which supports the suggestion by Ferraro et al. (2009) that 
Terzan 5 may be a remnant primordial building block of the Galactic bulge. 

The bulge globular clusters preserve chemical information about the early 
proto--bulge environment.  Therefore, understanding the connection between the 
bulge clusters and the broader bulge/bar system is necessary for interpreting
the complex chemodynamical stellar populations that currently reside in the 
inner Galaxy.  However, only a handful of bulge clusters have been examined in 
detail using high resolution spectroscopy.  Contamination from the bulge field 
star population and strong differential reddening complicate both integrated 
light and color--magnitude diagram analyses of bulge globular clusters.  
Therefore, we provide here new composition and kinematic analyses of the 
moderately metal--rich bulge globular clusters NGC 6342 and NGC 6366, based on 
high resolution optical spectra obtained with the MMT--Hectochelle and 
WIYN--Hydra spectrographs.  Low resolution spectroscopic analyses and 
color--magnitude diagram fits estimate that both clusters have 
[Fe/H] $\sim$ --0.6 (e.g., Da Costa \& Seitzer 1989; Heitsch \& Richtler 1999; 
Valenti et al. 2004; Origlia et al. 2005a; Saviane et al. 2012; Campos et al. 
2013), but only NGC 6342 has had some of its stars (4; Origlia et al. 2005a) 
examined via high resolution spectroscopy.  Although little is known about the 
chemical composition of NGC 6366, the cluster is particularly noteworthy 
because it has an unusual bottom--light mass function (Paust et al. 2009).  
NGC 6366 also has a very low velocity dispersion of $\sim$ 1 km s$^{\rm -1}$ 
(Da Costa \& Seitzer 1989; Rutledge et al. 1997), and may have experienced 
significant tidal stripping and mass loss (Paust et al. 2009).

In this paper, we examine the light odd--Z, $\alpha$, Fe--peak, and 
heavy element abundance patterns of NGC 6342 and NGC 6366 to compare with 
those of similar metallicity bulge cluster and field stars.  The addition of 
these new data to the literature will: allow for an investigation of the 
chemical similarities and differences between bulge cluster and field stars, 
define the RGB sequence of each cluster, permit further investigation into 
whether or not the typical light element abundance variations found in 
nearly all metal--poor clusters extend also to metal--rich bulge clusters, and 
help constrain the contribution of dissolved globular clusters to the bulge 
field.

\section{OBSERVATIONS, TARGET SELECTION, AND DATA REDUCTION}

\subsection{Observations and Target Selection}

The spectra for this project were obtained using the Hectochelle 
(Szentgyorgyi et al. 2011) and Hydra (Bershady et al. 2008; Knezek et al. 
2010) multi--fiber bench spectrographs mounted on the MMT 6.5m and WIYN 3.5m 
telescopes, respectively.  NGC 6342 was observed with Hectochelle on 18 June
2014 and also with Hydra on 27 June 2013.  However, NGC 6366 was only observed
with Hydra on 18 May 2012.  The Hectochelle observations consisted of a
2400 and 2065 sec exposure set using the 110 line mm$^{\rm -1}$ Echelle 
grating, the ``CJ26" filter, and 2$\times$1 (spatial$\times$dispersion) binning
to achieve a resolving power of 
R $\equiv$ $\lambda$/$\Delta$$\lambda$ $\approx$ 38,000.  Similarly, the Hydra
observations consisted of 3$\times$3600 sec exposures with the bench configured
to use the 316 line mm$^{\rm -1}$ Echelle grating, the X18 filter, the red 
camera and fibers, and 2$\times$1 binning to achieve a resolving power of
R $\approx$ 18,000.  The spectra spanned approximately 6140--6310 \AA\ for the 
Hectochelle data and 6080--6390 \AA\ for the Hydra data.

The target stars for both clusters were selected using photometry and 
coordinates available through the Two Micron All Sky Survey (2MASS; Skrutskie 
et al. 2006) database.  Since both clusters are located at relatively low
Galactic latitudes near the bulge, the fiducial RGB sequences for each cluster
are hidden by the significant stellar crowding.  Identifying the cluster RGB 
sequences is further complicated because the cluster stars and a large fraction
of the outer bulge field stars share similar metallicities.  Both clusters 
also are affected by significant reddening with NGC 6342 having 
E(B--V) $\approx$ 0.60 and $\Delta$E(B--V) $\approx$ 0.40 (Heitsch \& Richtler 
1999; Valenti et al. 2004; Alonso--Garc{\'{\i}}a et al. 2012) and NGC 6366
having E(B--V) $\approx$ 0.70 and $\Delta$E(B--V) $\approx$ 0.05 (Alonso et al.
1997; Sarajedini et al. 2007; Paust et al. 2009; Dotter et al. 2010; Campos et
al. 2013).  The combination of these effects makes it difficult to know
\emph{a priori} which stars are true cluster members.

\begin{figure}
\epsscale{1.00}
\plotone{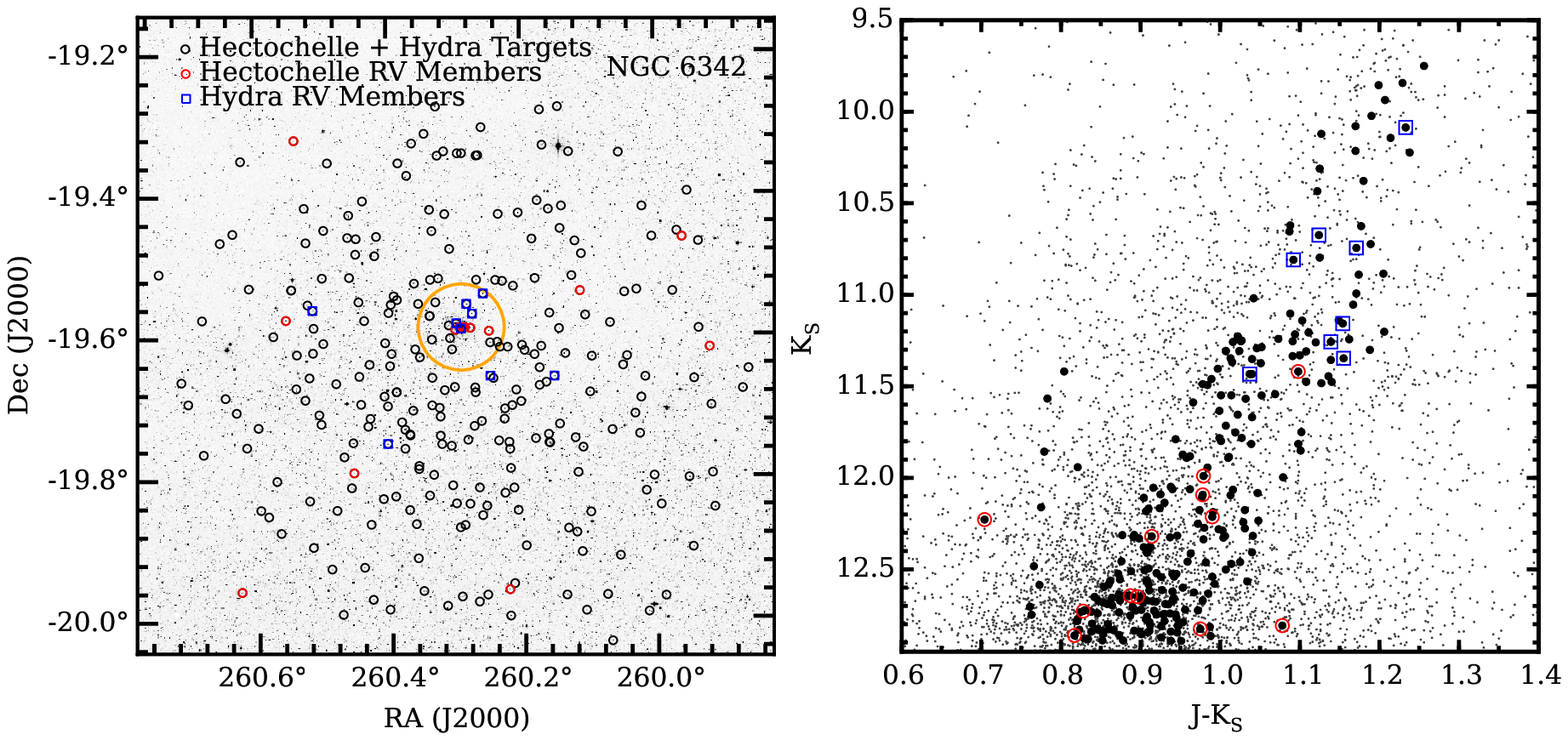}
\caption{\emph{left:} A Second Palomar Observatory Sky Survey r--band image of
NGC 6342 is shown, and the open black circles designate the coordinates of all 
stars observed with MMT--Hectochelle and WIYN--Hydra.  The open red circles 
(Hectochelle) and open blue boxes (Hydra) indicate the stars that have radial 
velocities consistent with cluster membership.  The solid orange contour line 
illustrates a distance of 5 times the cluster's 0.73$\arcmin$ half--light 
radius (Harris 1996; 2010 version).  \emph{right:} A K$_{\rm S}$ versus 
J--K$_{\rm S}$ color--magnitude diagram is shown for targets within 30$\arcmin$
of NGC 6342.  The small filled grey circles are all of the stars from the 
2MASS database (Skrutskie et al. 2006).  The red and blue symbols are the same 
as in the left panel, and the filled black circles indicate all stars that were
observed with both instruments.  Note that the two stars 
2MASS 17220959--1919193 and 2MASS 17223024--1957315 have velocities that may 
be consistent with cluster membership but are significantly bluer and redder 
than the main giant branch.}
\label{f1}
\end{figure}

\begin{figure}
\epsscale{1.00}
\plotone{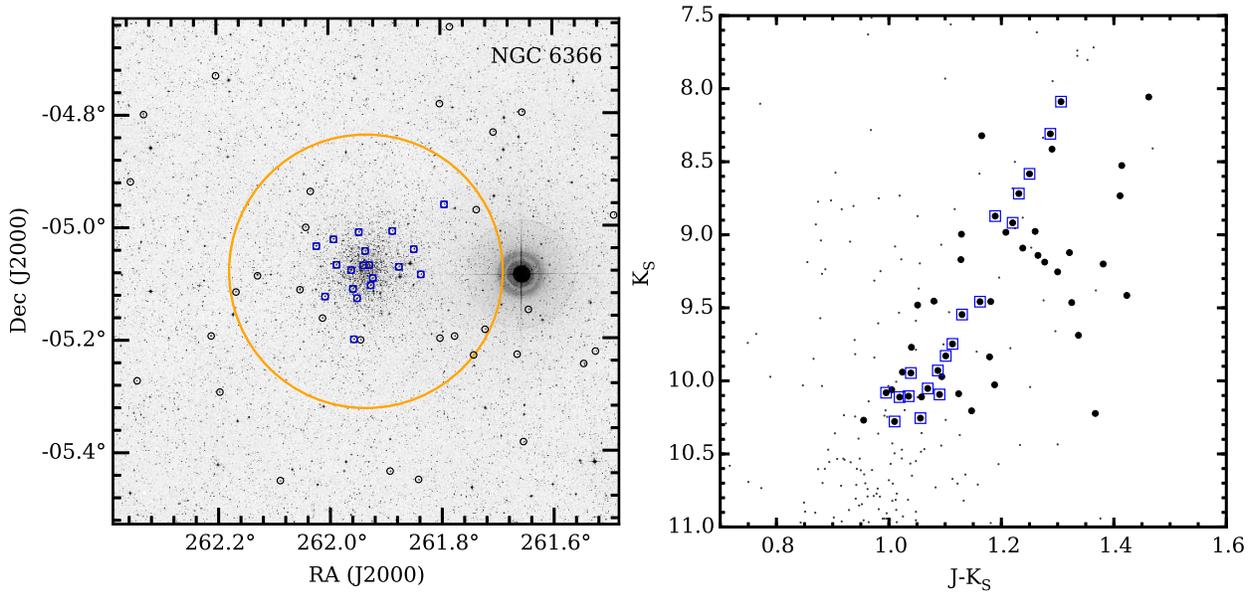}
\caption{\emph{left:} Similar to Figure \ref{f1}, the open black circles
show the coordinates for all stars near NGC 6366 that were observed with the
WIYN--Hydra instrument.  The targets that have radial velocities
consistent with cluster membership are designated by open blue boxes.
The solid orange contour line illustrates a distances of 5 times the cluster's 
2.92$\arcmin$ half--light radius (Harris 1996; 2010 version).
\emph{right:} A K$_{\rm S}$ versus J--K$_{\rm S}$ color--magnitude diagram is 
shown for targets within 30$\arcmin$ of NGC 6366.  The small filled grey 
circles are all stars from the 2MASS database (Skrutskie et al. 2006).  The 
black and blue symbols are the same as those in Figure \ref{f1}.}
\label{f2}
\end{figure}

Therefore, we repeated the selection procedure used in Johnson et al. (2015)
to identify cluster members in the bulge globular cluster NGC 6273.  Briefly,
we assumed the cluster RGB sequences could be reasonably well traced using only
stars within 2$\arcmin$ of each cluster's core.  The selection region was then 
broadened to include the effects of differential reddening, and stars were 
prioritized in the fiber assignment codes according to the distance between a
star and the cluster core.  A total of 216 fibers were placed on targets with
Hectochelle for NGC 6342 and 51 fibers were placed on NGC 6342 targets with
Hydra.  Similarly, 51 Hydra fibers were also placed on targets for NGC 6366.
The final selection of targets is illustrated in Figure \ref{f1} for NGC 6342 
and Figure \ref{f2} for NGC 6366.  The significantly larger differential 
reddening in NGC 6342 is clearly evident when comparing Figures \ref{f1} and 
\ref{f2}, and is largely the cause of the significantly lower percentage of
radial velocity members found in NGC 6342 (8$\%$) versus NGC 6366 (37$\%$; see
also Section 4).  The strict fiber--to--fiber distance restrictions of 
Hectochelle and Hydra contribute further to the low membership percentages
because the magnetic buttons cannot be packed efficiently near the cluster
cores, where the field contamination is at a minimum.  The star 
identifications, J2000 coordinates, 2MASS photometry, and radial velocity 
measurements for all member and non--member stars are provided in Tables 1--2.

\subsection{Data Reduction}

The data reduction process for both the Hectochelle and Hydra spectra was 
carried out using standard IRAF\footnote{IRAF is distributed by the National
Optical Astronomy Observatory, which is operated by the Association of
Universities for Research in Astronomy, Inc., under cooperative agreement with
the National Science Foundation.} tasks.  The raw spectra were bias subtracted 
and trimmed before the more specialized tasks of aperture identification and
tracing, scattered light removal, flat--field correction, ThAr wavelength
calibration, cosmic--ray removal, spectrum extraction, and sky subtraction 
were carried out.  Note that the sky subtraction was performed using 
simultaneous sky spectra obtained with fibers placed on ``blank" sky regions
in the Hectochelle and Hydra fields--of--view.  While a majority of the Hydra 
data reduction was performed using the IRAF task \emph{dohydra}, 
most of the Hectochelle reduction procedures were carried out by a dedicated 
pipeline maintained at the Smithsonian Astrophysical Observatory's Telescope 
Data Center.  

The final data reduction steps of telluric subtraction, continuum fitting, and 
spectrum combining were carried out using the IRAF tasks \emph{telluric}, 
\emph{continuum}, and \emph{scombine} outside the general pipelines.  The final
combined spectra yielded signal--to--noise (S/N) ratios of approximately 
50--100 per resolution element.  Due to higher extinction, worse observing 
conditions, and shorter exposures (for Hectochelle), the NGC 6342 data tended 
to have lower S/N than the NGC 6366 data.  However, we only measured abundances
in stars with the highest quality spectra, and for which we could measure 
$>$10 Fe I lines (see Table 3).

\section{DATA ANALYSIS}

\subsection{Model Atmospheres}

The model atmosphere parameters effective temperature (T$_{\rm eff}$), 
metallicity ([M/H]), and microturbulence ($\xi$$_{\rm mic.}$) were determined 
using spectroscopic methods.  Specifically, temperatures were set by removing
any trends in plots of log $\epsilon$(Fe I) abundance versus excitation 
potential, and microturbulence values were set by removing any trends in plots 
of log $\epsilon$(Fe I) abundance versus reduced equivalent width\footnote{The
reduced equivalent width is defined as log(EW/$\lambda$).} (EW).  The model 
atmosphere metallicities were set to the measured [Fe/H] values.  

\begin{figure}
\epsscale{1.00}
\plotone{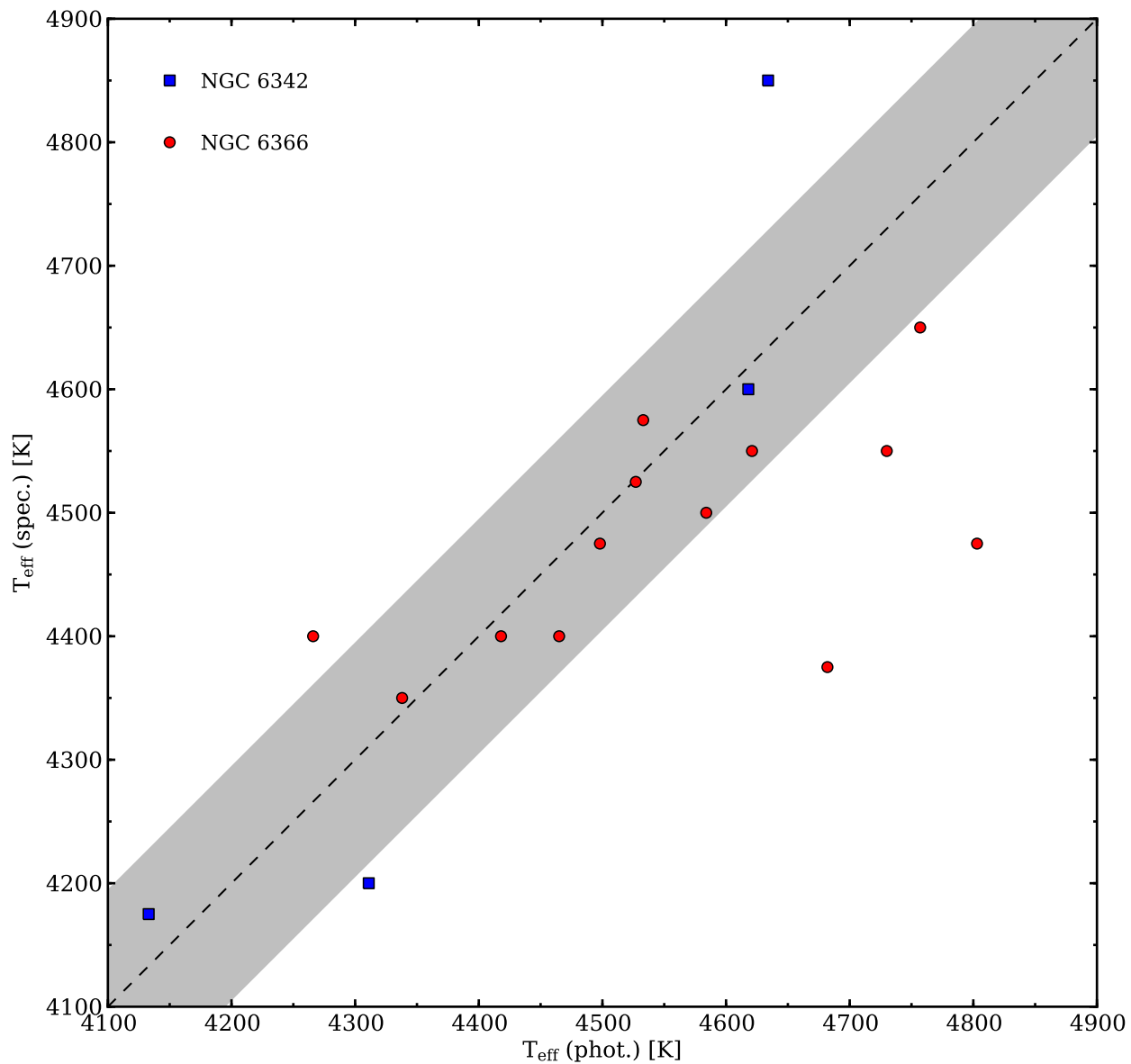}
\caption{A comparison between the effective temperatures derived by enforcing
excitation equilibrium of Fe I (T$_{\rm eff}$ spec.) and using the 
J--K$_{\rm S}$ color--temperature relation (T$_{\rm eff}$ phot.) from 
Gonz{\'a}lez Hern{\'a}ndez \& Bonifacio (2009) for NGC 6342 (filled red 
circles) and NGC 6366 (filled blue boxes).  The dashed black line indicates
perfect agreement, and the shaded region illustrates the 1$\sigma$ temperature
uncertainty from Gonz{\'a}lez Hern{\'a}ndez \& Bonifacio (2009).}
\label{f3}
\end{figure}

In Figure \ref{f3}, we compare the spectroscopic T$_{\rm eff}$ values derived 
using excitation equilibrium with the J--K$_{\rm S}$ color--temperature 
relation provided by Gonz{\'a}lez Hern{\'a}ndez \& Bonifacio (2009), assuming
E(B--V) = 0.57 for NGC 6342 (Valenti et al. 2004) and E(B--V) = 0.70 for 
NGC 6366 (Alonso et al. 1997).  Despite the presence of significant 
differential reddening in both clusters, the two temperature scales are 
well--correlated.  We find an average offset ($\Delta$T$_{\rm eff}$), in the 
sense of photometric T$_{\rm eff}$ minus spectroscopic T$_{\rm eff}$, to be 
$\Delta$T$_{\rm eff}$ = --32 K ($\sigma$ = 138 K) for NGC 6342 and 
$\Delta$T$_{\rm eff}$ = $+$77 K ($\sigma$ = 131 K) for NGC 6366.  Although the 
agreement is comparable to the 94 K standard deviation of the 
color--temperature relation (see Gonz{\'a}lez Hern{\'a}ndez \& Bonifacio 2009; 
their Table 5), the removal of one significant outlier in NGC 6342 and two 
significant outliers in NGC 6366 decreases the offsets to 
$\Delta$T$_{\rm eff}$ = $+$32 K ($\sigma$ = 77 K) and 
$\Delta$T$_{\rm eff}$ = $+$33 K ($\sigma$ = 83 K), respectively.

Since the data span a limited wavelength range, we did not constrain surface 
gravities (log(g)) by setting ionization equilibrium between neutral and singly
ionized species (e.g., Fe I/II).  Instead, we estimated surface gravities using
isochrones available through the Dartmouth Stellar Evolution database (Dotter 
et al. 2008).  We used the online interpolator\footnote{The Dartmouth Stellar 
Evolution database online interpolator can be accessed at: 
http://stellar.dartmouth.edu/models/isolf$\_$new.html.} to obtain a surface 
gravity value for each star, given its spectroscopically determined 
temperature.  We assumed a standard helium mass fraction, 
[$\alpha$/Fe] = $+$0.4 (Origlia et al. 2005a; see also Section 5), and an age 
of 11 Gyr for both clusters (e.g., Campos et al. 2013; VandenBerg et al. 2013).

Each model atmosphere was calculated by interpolating within the 
$\alpha$--enhanced ATLAS9 grid (Castelli \& Kurucz 2004)\footnote{The model 
atmosphere grid can be accessed at: 
http://wwwuser.oats.inaf.it/castelli/grids.html.}.  The final values were 
determined by simultaneously solving for temperature, metallicity, and 
microturbulence and iteratively redetermining surface gravity via the isochrone
interpolator mentioned above.  The final adopted parameters for NGC 6342 
(4 stars) and NGC 6366 (13 stars) are provided in Table 3.

\subsection{Abundance Analysis}

The abundance analysis for this work closely follows that described in Johnson
et al. (2014).  Briefly, the abundances of Si, Ca, Cr, Fe, and Ni were 
calculated using the \emph{abfind} driver of the LTE line analysis code 
MOOG (Sneden 1973; 2014 version).  Similarly, the abundances of O, Na, Mg, and 
La were determined using the \emph{synth} driver of MOOG to minimize 
differences between the observed and synthetic spectra.  All EWs were 
measured using the semi--automated code outlined in Johnson et al. (2014) that 
fits single or multiple Gaussian profiles to isolated and blended absorption
lines.  

Since previous estimates indicate that both clusters should have 
[Fe/H] $\sim$ --0.6 (see Section 1), we performed a differential abundance
analysis relative to the giant star Arcturus ([Fe/H] $\approx$ --0.5; e.g.,
Ram{\'{\i}}rez \& Allende Prieto 2011).  The line list provided in 
Table 4 uses the same log(gf) values, solar reference abundances, and Arcturus 
reference abundances as those in Johnson et al. (2014), but we added a few 
additional Fe I lines due to differences in wavelength coverage between the 
FLAMES, Hectochelle, and Hydra data.  The only other exception to the Johnson 
et al. (2014) line list was our inclusion of the La atomic data from Lawler et 
al. (2001), which takes into account hyperfine structure introduced by the 
\iso{139}{La} isotope.

For the abundances determined by spectrum synthesis, we included atomic lines
from the Kurucz database\footnote{The most up--to--date line lists can be 
accessed at: http://kurucz.harvard.edu/linelists/gfnew/.} and CN molecular
lines from Sneden et al. (2014).  The original log(gf) values were manually 
adjusted until the synthetic spectrum matched the observed spectrum of the 
Arcturus atlas (Hinkle et al. 2000).  We adopted the model atmosphere 
parameters and Arcturus abundances listed in Table 4 and Ram{\'{\i}}rez \& 
Allende Prieto (2011) for C, N, O, Na, Mg, and La.  In order to account for
contributions from CN lines, we initially assumed the stars were well--mixed
with [C/Fe] = --0.30, [N/Fe] = $+$0.50, and \iso{12}{C}/\iso{13}{C} = 4, 
which are typical values for evolved RGB stars in the bulge field and clusters 
(e.g., Origlia et al. 2005a; Mel{\'e}ndez et al. 2008; Ryde et al. 2010).
Since C, N, and O are integral parts of the molecular equilibrium calculations,
we first fit the 6300 [O I] line and nearby CN features to define
the oxygen and C$+$N abundances for each star.  The nitrogen abundance was
treated as a free parameter in order to fit the CN line profiles.  For the two 
stars in NGC 6342 that we were not able to measure [O/Fe], we assumed 
[O/Fe] = $+$0.60, the average oxygen abundance for the other two NGC 6342 
stars in our sample, and set [C/Fe] = --0.30 and [N/Fe] = $+$0.50.  

While the 6154/6160 \AA\ Na I and 6262 \AA\ La II lines are relatively 
uncontaminated, the 6318--6319 \AA\ Mg I triplet lines are moderately affected 
by a broad Ca I auto--ionization feature.  In order to correct for this effect,
we adjusted the log $\epsilon$(Ca) abundance until the shape of the synthetic 
and observed spectra matched in the nearby continuum windows (e.g., see Figure 
6 of Johnson et al. 2014).  The log $\epsilon$(Ca) abundance that best 
reproduced the auto--ionization feature tended to be $\sim$0.3--0.4 dex less
than the average log $\epsilon$(Ca) abundance measured from the individual 
atomic lines.
 
\subsubsection{Internal Abundance Uncertainties}

The largest sources of internal abundance uncertainties are typically related
to the derivation of the stellar model atmosphere parameters.  Additional
sources, such as line blending, continuum placement, atomic parameters, and
visual profile fitting, tend to have only a small effect on the final 
abundances derived from moderately high S/N and resolution spectra.  The line 
profile fitting code used for this project takes into account a spectrum's S/N 
and estimates the uncertainty range in continuum placement.  The 
continuum uncertainty is then propagated through the profile fitting procedure, 
and the code generates an EW error estimate for every line.  The average EW 
uncertainty ranges from approximately 10$\%$ for a line of 20 m\AA\ 
to 2$\%$ for a line of 150 m\AA.  These uncertainties translate into abundance 
errors of $\sim$0.02--0.05 dex, which are comparable to the typical standard
errors of the mean derived in our analysis ($\sim$0.03 dex on average).  
Therefore, our final internal uncertainty calculations provided in Table 5 
include the error of the mean for each element as a tracer of the random 
measurement error.

In order to examine the internal sensitivity of the derived log $\epsilon$(X)
abundances to changes in the model atmosphere parameters, we calculated the 
abundance differences between the ``best--fit" model and those with each 
parameter adjusted within its estimated uncertainty range.  A temperature 
uncertainty of 75 K was adopted based on a comparison of the spectroscopic and 
photometric T$_{\rm eff}$ values shown in Figure \ref{f3}, where the 1$\sigma$ 
star--to--star deviation, after removing three extreme outliers, is 79 K.  
Since the surface gravity values were determined from the isochrones 
described in Section 3.1, we estimated that the interpolation uncertainty
for log(g) was 0.10 cgs, assuming $\Delta$T$_{\rm eff}$ = 75 K and 
$\Delta$age = 1 Gyr.  The overall metallicity uncertainty was estimated to
be 0.10 dex based on the star--to--star dispersion in our derived [Fe/H] 
values for both clusters.  Finally, we estimated that the microturbulence
velocity uncertainty was 0.10 km s$^{\rm -1}$, based on an examination of 
the line--to--line scatter in plots of log $\epsilon$(Fe I) abundance 
versus reduced EW.  The final abundance uncertainties, including the 
measurement errors described above, were added in quadrature to produce the
final uncertainty values listed in Table 5.  Note that the [X/Fe] ratios take
into account both the errors in [Fe/H] and [X/H] for each element.  The typical
errors range from $\sim$0.05--0.10 dex.

\section{RADIAL VELOCITY MEASUREMENTS AND CLUSTER MEMBERSHIP}

The radial velocities for both NGC 6342 and NGC 6366 were measured using 
the IRAF task \emph{fxcor}, which cross correlated the observed spectra with a 
convolved and rebinned version of the Arcturus atlas (Hinkle et al. 2000) that
matches the resolution and sampling of the Hectochelle and Hydra spectra.
We avoided contamination due to any residual telluric lines by only using
the 6120--6275 \AA\ region for the cross correlation.  The heliocentric 
corrections were determined using the information in the image headers and the
IRAF task \emph{rvcor}.  Since there were two exposures for the Hectochelle
data and three exposures each for the Hydra data, we measured the heliocentric
radial velocity (RV$_{\rm helio.}$) in each exposure and treated these values
as independent measurements.  The standard deviation of these measurements for
each star are listed in Tables 1--2 as the RV$_{\rm helio.}$ errors.  The 
average RV$_{\rm helio.}$ error for the entire sample is 0.38 km s$^{\rm -1}$
($\sigma$ = 0.31 km s$^{\rm -1}$); however, the Hectochelle data have a 
higher resolution and thus a lower average error of 0.35 km s$^{\rm -1}$ 
($\sigma$ = 0.20 km s$^{\rm -1}$) compared to the 0.44 km s$^{\rm -1}$
($\sigma$ = 0.45 km s$^{\rm -1}$) average error for the Hydra data.  

Three stars in the field of NGC 6342 (2MASS 17221115--1931581, 
17213359--1940422, and 17212139--1934169) were observed with both Hectochelle 
and Hydra (see also Table 1), and the independent RV$_{\rm helio.}$ 
measurements permit a rough estimate of the zero point offset between the two 
observing runs.  The RV$_{\rm helio.}$ differences, in the sense of Hectochelle
minus Hydra, are $+$0.18 km s$^{\rm -1}$, $+$6.10 km s$^{\rm -1}$, and 
$+$1.11 km s$^{\rm -1}$ for 2MASS 17221115--1931581, 17213359--1940422, and 
17212139--1934169, respectively.  If we neglect the large RV$_{\rm helio.}$ 
difference for 2MASS 17213359--1940422, which could be a velocity variable or
binary star, the average zero point offset between Hectochelle and Hydra is 
$+$0.65 km s$^{\rm -1}$.

Given the significant stellar crowding, differential reddening, and similar
metallicities of bulge field stars to the NGC 6342 and NGC 6366 cluster stars, 
radial velocities are useful indicators of cluster membership.  Previous
analyses measured heliocentric radial velocities of NGC 6342
that range from $\sim$$+$114 to $+$118 km s$^{\rm -1}$ with velocity
dispersions that range from $\sim$5--8 km s$^{\rm -1}$ (Dubath et al. 1997; 
Origlia et al. 2005; Casetti--Dinescu et al. 2010).  Similarly, the 
heliocentric radial velocity estimates for NGC 6366 range from --122.6 
to --123.2 km s$^{\rm -1}$, but the dispersion is estimated to be only 
$\sim$1 km s$^{\rm -1}$ (Da Costa \& Seitzer 1989; Rutledge et al. 1997).  We 
find in agreement with past work that the average RV$_{\rm helio.}$ of 
NGC 6342 is $+$112.5 km s$^{\rm -1}$ ($\sigma$ = 8.6 km s$^{\rm -1}$) and 
that of NGC 6366 is --122.3 km s$^{\rm -1}$ ($\sigma$ = 1.5 km s$^{\rm -1}$).
 
\begin{figure}
\epsscale{1.00}
\plotone{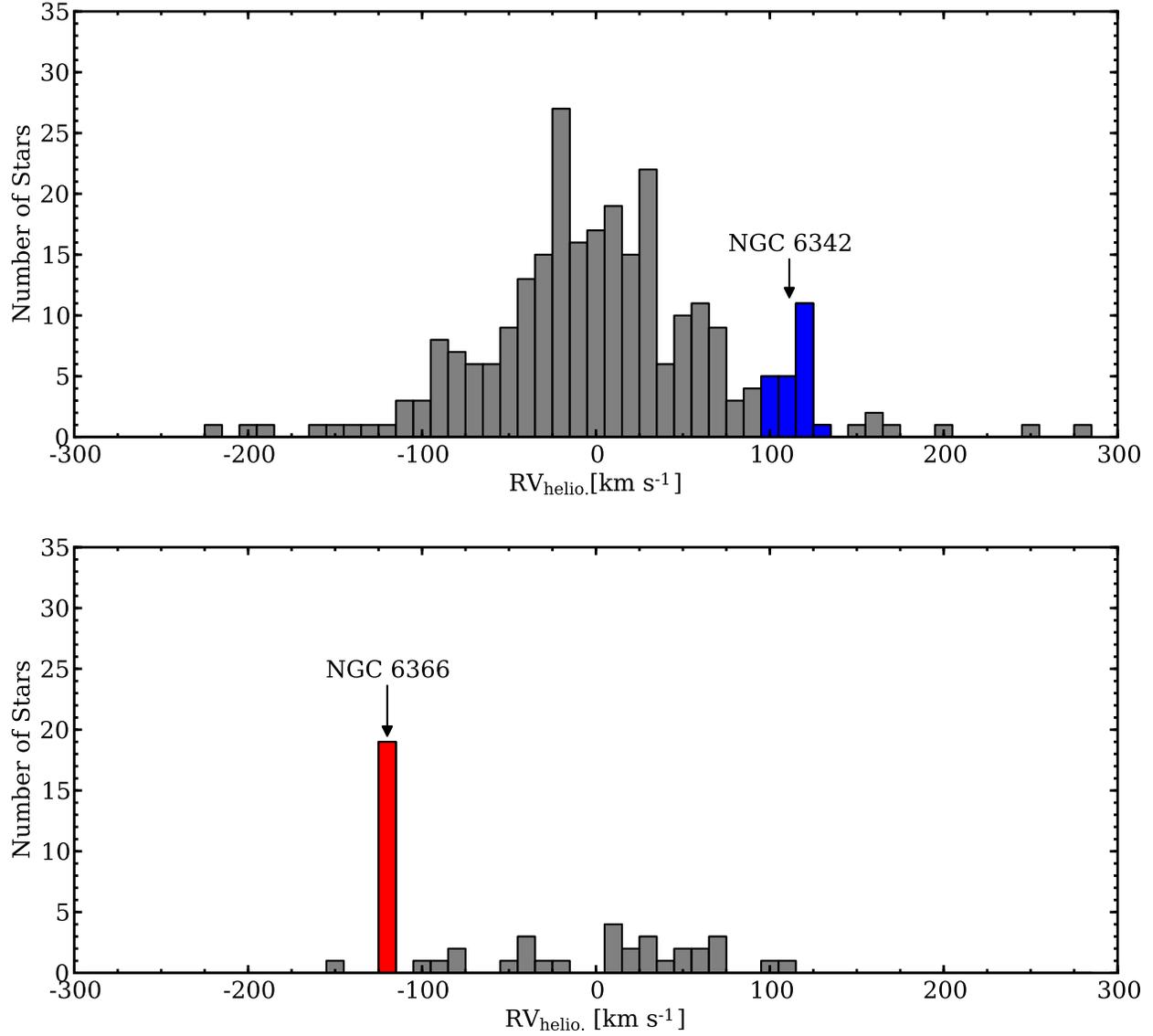}
\caption{The heliocentric radial velocity distributions for the fields near 
NGC 6342 (top) and NGC 6366 (bottom) are shown with bin sizes of 10 km 
s$^{\rm -1}$.  The high probability members for each cluster are highlighted
in blue for NGC 6342 and red for NGC 6366.}
\label{f4}
\end{figure}

Although the broad radial velocity distribution of the bulge field stars 
is consistent with observations from recent large sample studies (e.g., Kunder 
et al. 2012; Ness et al. 2013; Zoccali et al. 2014), the field contamination
makes assigning membership more difficult in the NGC 6342 field than the 
NGC 6366 field (see Figure \ref{f4}).  Therefore, we have only included stars
with RV$_{\rm helio.}$ values between 95--130 km s$^{\rm -1}$ ($\sim$2$\sigma$)
as possible cluster members for NGC 6342.  Nevertheless, as mentioned in 
Section 2.1, only 8$\%$ and 37$\%$ of the observed targets are likely radial 
velocity members of NGC 6342 and NGC 6366, respectively.  However, a high 
contamination rate of field stars is typical for bulge globular cluster 
observations that extend far beyond the cluster core (e.g., Gratton et al. 
2007).

\section{RESULTS AND DISCUSSION}

\subsection{Comparing NGC 6342 and NGC 6366}

As mentioned in Section 1, neither NGC 6342 nor NGC 6366 has been extensively
studied with high resolution spectroscopy.  Origlia et al. (2005a) obtained
high resolution infrared spectra of four RGB stars in NGC 6342 and found the 
cluster to be only moderately metal--poor ($\langle$[Fe/H]$\rangle$ = --0.60), 
enhanced in all $\alpha$--elements ($\langle$[$\alpha$/Fe]$\rangle$ = $+$0.34), 
and exhibit a small star--to--star dispersion in [O/Fe] ($\Delta$[O/Fe] = 0.04 
dex).  NGC 6366 has never been analyzed with high resolution spectroscopy, but 
photometric analyses suggest that the cluster may host at least two 
populations with different light element chemistry (Piotto et al. 2015; their 
Figure 14).  Photometric and low/moderate resolution spectroscopy further 
indicate that NGC 6366 is comparable in metallicity to NGC 6342, with estimates
ranging from [Fe/H] = --0.85 to --0.54 (Johnson et al. 1982; Da Costa \& 
Seitzer 1989; Da Costa \& Armandroff 1995; Alonso et al. 1997; Dotter et al. 
2010; Saviane et al. 2012; Campos et al. 2013).

The chemical composition results presented here, including 4 RGB stars for 
NGC 6342 and 13 RGB stars for NGC 6366, are in general agreement with previous 
work.  For NGC 6342, we find an average [Fe/H] = --0.53 ($\sigma$ = 0.11), 
significantly enhanced [O/Fe] ($\langle$[O/Fe]$\rangle$ = $+$0.61) with 
smaller enhancements for the heavier $\alpha$--elements 
($\langle$[$\alpha$/Fe]$\rangle$ = $+$0.33), and a moderate spread in [O/Fe] 
($\Delta$[O/Fe] = 0.32).  For NGC 6366, we find an average [Fe/H] = --0.55 
($\sigma$ = 0.09), a slightly lower average [O/Fe] = $+$0.51 with similar 
enhancements for the heavier $\alpha$--elements 
($\langle$[$\alpha$/Fe]$\rangle$ = $+$0.29), and a slightly larger spread in
[O/Fe] ($\Delta$[O/Fe] = 0.40).  Both clusters also exhibit moderate 
enhancements and dispersions of [Na/Fe], and possibly [La/Fe], but 
the Fe--peak elements in both clusters mostly track iron\footnote{The 0.14 dex
dispersion in [Cr/Fe] for NGC 6366 is likely due to increased measurement 
errors and the availability of only one weak line}.  The data indicate that
NGC 6366 may have [Cr/Fe] and [Ni/Fe] abundances that are $\sim$ 0.1 dex 
higher than those of NGC 6342, but a larger sample of Fe--peak element
measurements in these two clusters is needed in order to verify that this 
difference is real.  We note that NGC 6342 and NGC 6366 have similar 
differences between their maximum and minimum [La/Fe] abundances 
($\Delta$[La/Fe] $\sim$ 0.35 dex), but different dispersions.  Therefore, the 
larger dispersion in NGC 6342 may be a result of the small sample size 
(4 stars); however, we cannot rule out that the smaller [La/Fe] dispersion in 
NGC 6366 is a residual effect from enhanced tidal disruption (Paust et al. 
2009).  A summary of the average and dispersion values for every element in 
each star per cluster is provided in Table 6.

\subsection{NGC 6342, NGC 6366, and the Bulge Globular Cluster System}

\begin{figure}
\epsscale{1.00}
\plotone{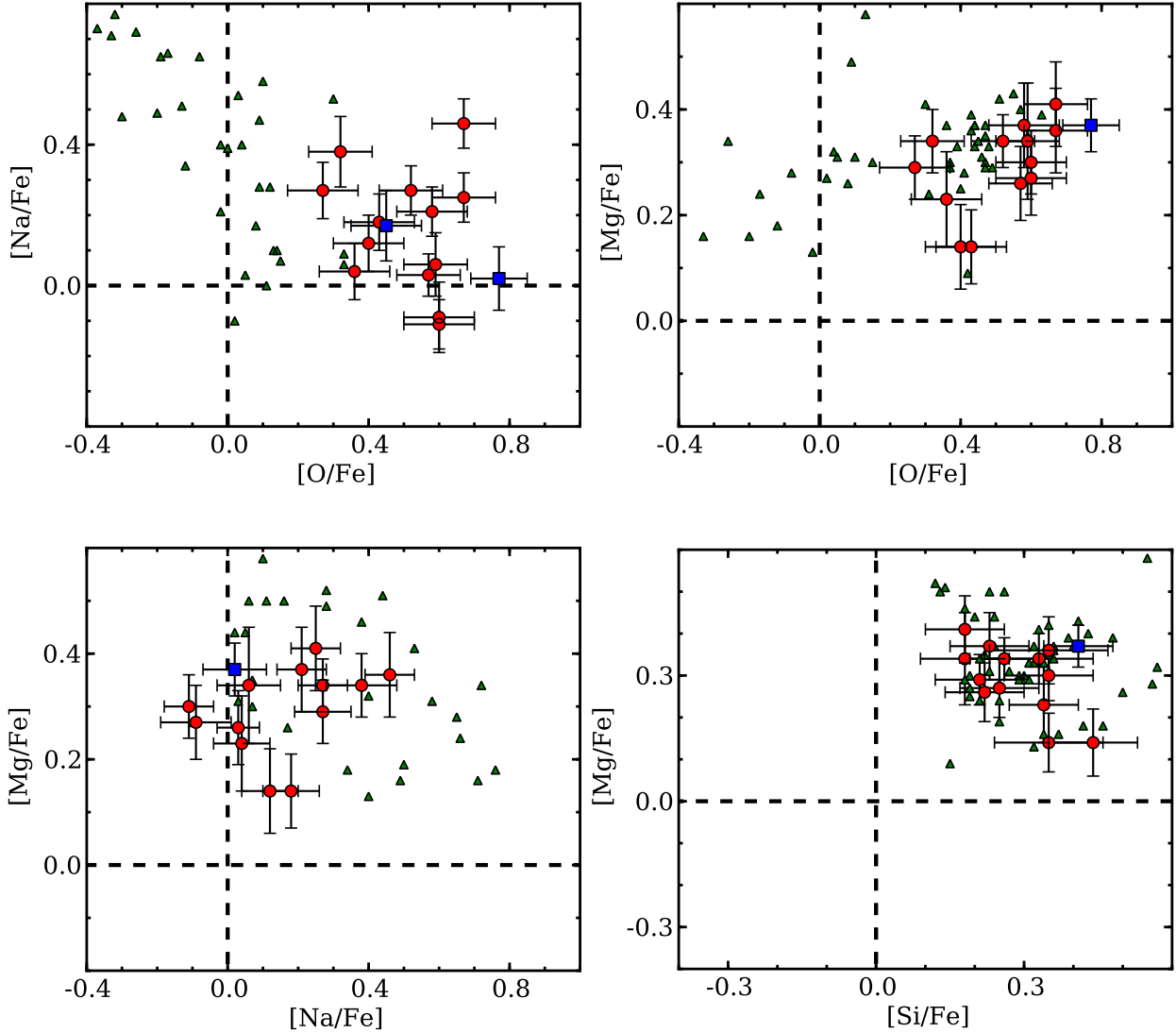}
\caption{Similar to Figure 11 of Carretta (2015), plots of the O--Na, O--Mg, 
Na--Mg, and Si--Mg distributions are shown for NGC 6342 and NGC 6366.  The
dashed black lines indicate the solar abundances ratios, and the 
symbols are the same as those in Figure \ref{f3}.  The filled green triangles 
in each panel indicate the abundance ratios for individual stars of the bulge 
globular clusters listed in Table 7.  For comparison purposes we have only 
included clusters that have [Fe/H] between --0.70 and --0.40, which are 
comparable to the metallicities of NGC 6342 and NGC 6366.}
\label{f5}
\end{figure}

In Figure \ref{f5}, we compare the light element abundance patterns of 
individual stars in NGC 6342 and NGC 6366 with those of similar metallicity
(--0.7 $\la$ [Fe/H] $\la$ --0.4) bulge globular clusters.  The relationships 
between the element ratio pairs shown in Figure \ref{f5} are often interpreted 
as being a result of high temperature (T $\ga$ 65$\times$10$^{\rm 6}$ K) 
proton--capture burning (e.g., Langer et al. 1993; Arnould et al. 1999; 
Prantzos et al. 2007; Ventura et al. 2012), and as a consequence many globular 
clusters exhibit O--Na and Mg--Si anti--correlations concurrent with O--Mg 
correlations (e.g., Yong et al. 2005; Carretta et al. 2009b; Johnson \& 
Pilachowski 2010; Cohen \& Kirby 2012; Carretta et al. 2014; Yong et al. 2014; 
Carretta 2015; Roederer \& Thompson 2015).  However, as can be seen in
Figure \ref{f5}, NGC 6366 only shows evidence supporting the existence of a 
moderately extended O--Na anti--correlation.  Although the NGC 6342 data 
overlap with the O--Na trend observed in NGC 6366 and other similar metallicity
bulge clusters, we were only able to measure [O/Fe] and [Na/Fe] in two stars 
for NGC 6342 and therefore cannot comment further on the extent, or existence, 
of a true O--Na anti--correlation in this cluster.  

The NGC 6366 data do not show significant star--to--star abundance 
variations or correlations for [Mg/Fe] and [Si/Fe], and from these data we can 
speculate that the gas from which the cluster stars formed did not reach 
temperatures hot enough to burn significant amounts of Mg into Al nor Al into
Si.  Previous observations of metal--rich ([Fe/H] $\ga$ --1) globular clusters 
have noted similar trends of small [Mg/Fe], [Al/Fe], and/or [Si/Fe] 
star--to--star abundance dispersions, and several authors have suggested that 
the MgAl cycle may be less efficient at higher metallicities (e.g., Carretta et
al. 2004; Carretta et al. 2007; Carretta et al. 2009b; O'Connell et al. 2011; 
Cordero et al. 2014, 2015).  Although larger samples are still needed, the 
observations of NGC 6342 and NGC 6366 provided here support this idea.  The 
current work and literature data shown in Figure \ref{f5} suggest that most 
globular clusters with --0.7 $\la$ [Fe/H] $\la$ --0.4 have about the same 
[Mg/Fe] and [Si/Fe] distributions, regardless of the extent of their O--Na 
anti--correlations.

In Figure \ref{f6}, we plot the median [X/Fe] ratios and dispersions as a 
function of [Fe/H] for NGC 6342, NGC 6366, and several other metal--rich 
([Fe/H] $\ga$ --1) bulge globular clusters available in the literature.
We find that the median [X/Fe] ratios and dispersions of most elements in 
NGC 6342 and NGC 6366 are in good agreement with other similar metallicity 
globular clusters (see also Table 6).  Although the similar abundance trends 
of most elements heavier than Na in Figure \ref{f6} suggest a common formation 
environment, the different [La/Fe] distributions may indicate a more
heterogeneous formation process.  

For example, NGC 6342, NGC 6366, and NGC 6388 (Carretta et al. 2007) all 
exhibit similarly enhanced [La/Fe] ratios; however, HP--1 (Barbuy et al. 2006) 
and NGC 6553 (Alves--Brito et al. 2006) have [La/Fe] $\la$ 0.  Barbuy et al. 
(2009) noted a similar trend that stars in NGC 6522 were significantly more 
Ba/La--enhanced than those of HP--1 and NGC 6558, despite all three clusters 
sharing roughly similar [Fe/H], $\alpha$--element, and O--Na distributions.  
Additionally, Gratton et al. (2006) found that bulge clusters largely exhibited
similar abundance patterns, but that certain elements, such as Mn, may vary 
from cluster--to--cluster.  Therefore, unlike the $\alpha$--elements Mg, Si, 
and Ca, which appear similarly enhanced for nearly all metal--rich bulge 
clusters, the heavy elements may provide some discrimination regarding how, or 
where, inner Galaxy clusters formed.  The current data are insufficient to 
provide any definitive links between bulge clusters with similar heavy element 
abundances, but increased sample sizes will aid in the interpretation of the 
observed cluster--to--cluster heavy element abundance variations.

\subsection{Comparing Bulge Globular Clusters and Field Stars}

\begin{figure}
\epsscale{1.00}
\plotone{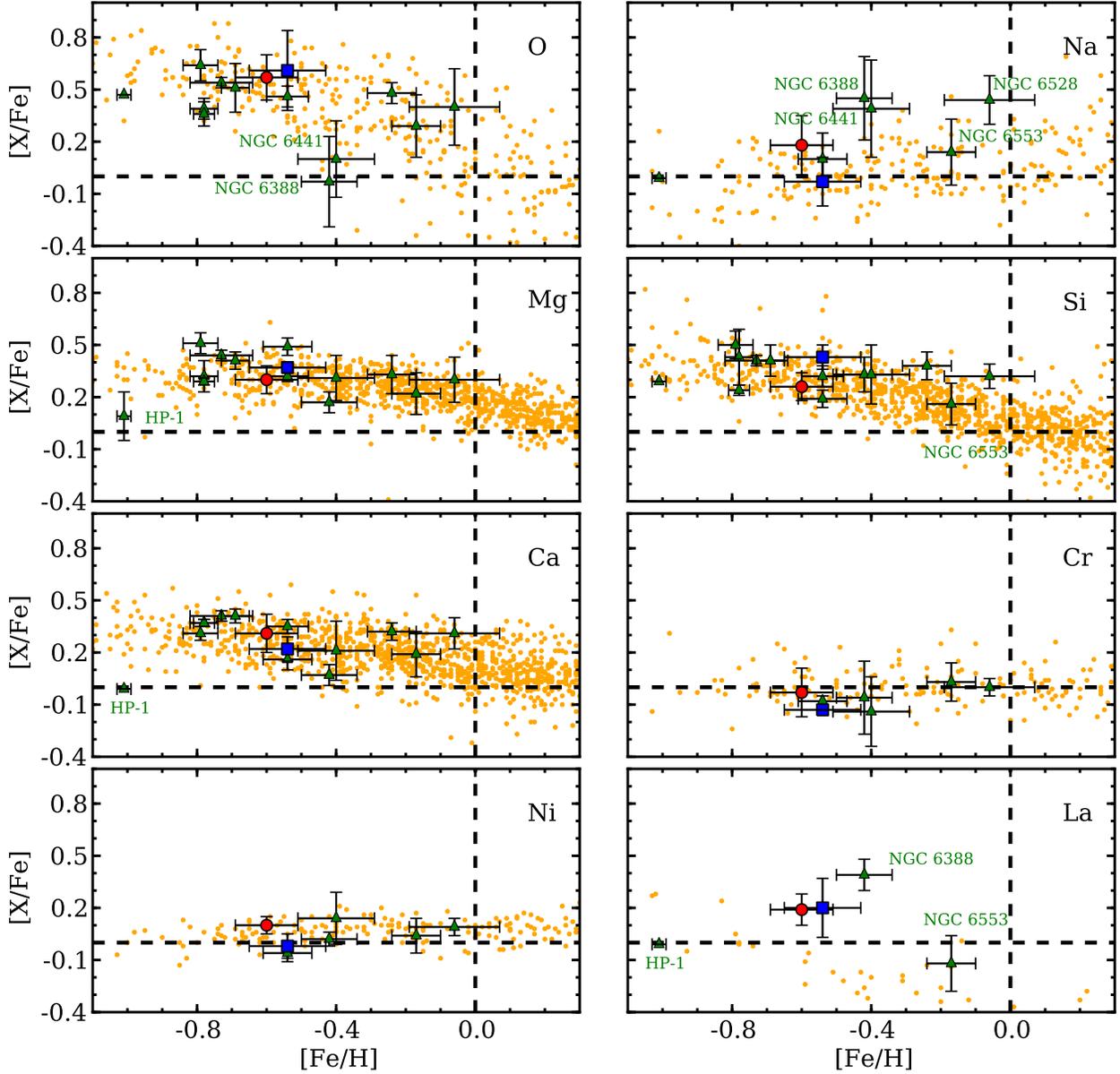}
\caption{The panels compare the [X/Fe] ratios of NGC 6342 (filled blue squares)
and NGC 6366 (filled red circles) as a function of [Fe/H] with bulge field star
(filled orange circles) and bulge globular cluster (filled green triangles) 
data from the literature.  The bulge globular cluster data are limited to
those having --1 $\la$ [Fe/H] $\la$ 0 and a Galactocentric distance 
(R$_{\rm GC}$) $\la$ 5 kpc.  For all clusters, the symbols indicate the median
abundance ratios and the error bars show the standard deviation.  The 
literature references are provided in Table 7.  Clusters of interest in each
panel are identified by name (see text for details).  For Terzan 5, only stars 
with [Fe/H] $<$ 0 have been included, and the metal--poor ([Fe/H] $\sim$ --0.8)
and intermediate metallicity ([Fe/H] $\sim$ --0.3) populations are shown as 
separate symbols.}
\label{f6}
\end{figure}

For stars with [Fe/H] $\la$ --0.4, previous investigations have largely found
that the bulge globular cluster and field star populations share similar
compositions (e.g., Carretta et al. 2001; Gratton et al. 2006; Gonzalez et al.
2011; Origlia et al. 2011; Johnson et al. 2014).  Figure \ref{f6} and Table 6
indicate that NGC 6342 and NGC 6366 continue this trend, and that the clusters 
only show significantly different patterns for [Na/Fe] (larger dispersion) and 
[La/Fe] (higher abundances).  The [O/Fe] dispersion may also be
larger for some clusters compared to the field stars; however, the larger 
measurement errors of [O/Fe], especially for field stars with uncertain 
gravities, make it more difficult to disentangle real scatter from 
measurement errors.

The larger [Na/Fe] dispersions in NGC 6342, NGC 6366, and other
metal--rich bulge clusters (e.g., Gratton et al. 2007; Carretta et al. 2007; 
Johnson et al. 2014), compared to similar metallicity bulge field stars, are 
not surprising and likely a result of self--enrichment processes occurring in 
the cluster environments.  On the other hand, the enhanced [La/Fe] abundances 
found in NGC 6342, NGC 6366, and NGC 6388 (Carretta et al. 2007; Worley \& 
Cottrell 2010) stars may be a reflection of the broader Galactic globular 
cluster trend.  In general, globular clusters tend to exhibit [Ba/Fe] and 
[La/Fe] ratios\footnote{Although Ba and La isotopes can be produced in both 
the r--process and s--process, at [Fe/H] $\ga$ --1, and also in the Solar 
System, these elements are predominantly produced by the s--process (e.g. see 
reviews by Busso et al. 1999; Sneden et al. 2008).} that increase between 
[Fe/H] = --2.5 and --1.5, and then remain enhanced at [Ba,La/Fe] $\sim$ $+$0.3 
to at least [Fe/H] = --0.5 (e.g., James et al. 2004; D'Orazi et al. 2010).  In 
contrast, bulge field stars show similar enhancements at [Fe/H] $\la$ --1, but 
the [Ba/Fe] and [La/Fe] ratios begin to decline at higher metallicity 
(e.g., McWilliam et al. 2010; Bensby et al. 2011; Johnson et al. 2011; 
Bensby et al. 2013; see also Figure \ref{f6}).  Therefore, the 
[Fe/H] $\sim$ --0.5 clusters NGC 6342, NGC 6366, and NGC 6388 exhibit [La/Fe] 
abundances that are more similar to those of bulge field stars with 
[Fe/H] $\la$ --1.  However, the different behavior of [Ba/Fe] and [La/Fe] 
between field and cluster stars of similar metallicity, at least between 
[Fe/H] = --1 and --0.4, suggests that the two populations experienced 
different formation and s--process enrichment histories.  

For metallicities higher than [Fe/H] $\sim$ --0.4, some composition differences
between bulge field and cluster stars may become more prevalent.  In particular,
Figure \ref{f6} indicates that [Mg/Fe], [Si/Fe], and [Ca/Fe] may remain 
enhanced to a higher metallicity in cluster stars than bulge field stars 
(e.g., see also Carretta et al. 2007).  Since the [$\alpha$/Fe] ratios of most 
clusters remain approximately constant and enhanced, at least up to 
[Fe/H] $\sim$ --0.1, the data suggest a stronger contribution from 
core--collapse supernovae (e.g., Tinsley 1979), and possibly also a more rapid 
formation timescale, for the inner Galaxy cluster population.  However, 
more globular cluster $\alpha$--element abundance measurements, especially
for clusters with [Fe/H] $\ga$ --0.3, are needed to definitively confirm that
the most metal--rich globular clusters have higher [$\alpha$/Fe] ratios than
similar metallicity bulge field stars.

Among the heavier elements discussed here (Cr, Ni, and La), only the [La/Fe] 
ratios show any evidence of discriminating bulge cluster stars from field 
stars at [Fe/H] $>$ --0.4.  The data from NGC 6553 (Alves--Brito et al. 
2006) shown in Figure \ref{f6} indicate marginally higher [La/Fe] abundances 
than the similar metallicity field stars.  Similarly, Carretta et al. (2001) 
found the near solar metallicity bulge cluster NGC 6528 to have an average 
[Ba/Fe] = $+$0.14, which is again marginally higher than the roughly solar 
[Ba/Fe] abundances found in microlensed bulge dwarf stars (Bensby et al. 2011;
Bensby et al. 2013)\footnote{The reader should note that if the Carretta et al.
(2001) data are compared with the Baade's window [Ba/Fe] abundances from
McWilliam \& Rich (1994), both populations exhibit similar [Ba/Fe] 
enhancements.}.  Although more data comparing the heavy element abundance
trends of metal--rich bulge cluster and field stars are needed, the small
samples available so far indicate that the stronger s--process signature found
in clusters near [Fe/H] = --0.5 may continue to at least solar metallicity.

Finally, a comparison between the chemical composition of bulge clusters with
--1 $\la$ [Fe/H] $\la$ 0 and similar metallicity field stars suggests that the
former population likely did not contribute a significant number of stars to
the latter population.  Second generation stars from clusters such as NGC 6388 
and NGC 6441 are strongly ruled out by their very low [O/Fe] and high [Na/Fe] 
ratios, and the clusters' first generation stars are incompatible with the 
field star composition based on their [La/Fe] abundances.  More typical 
clusters such as NGC 6342 and NGC 6366 are ruled out mostly by their large 
[Na/Fe] spreads and enhanced [La/Fe] abundances.  Similarly, clusters with 
[Fe/H] $\ga$ --0.4 are mostly ruled out by their higher [$\alpha$/Fe] 
abundances compared to the field stars.  Clusters such as Terzan 5 could have 
contributed stars to the bulge field (e.g., see Origlia et al. 2011; Origlia et
al. 2013), but it is not clear how many such objects exist nor is it clear 
that Terzan 5's composition will remain compatible with the field once its 
[Na/Fe] and heavy element abundances are measured.

\section{SUMMARY}

For this project, we used the MMT--Hectochelle and WIYN--Hydra spectrographs to
obtain high resolution spectra of 267 RGB stars and 51 RGB stars in the bulge 
globular clusters NGC 6342 and NGC 6366, respectively.  Cluster membership was
determined primarily through radial velocity measurements.  However, the 
significant reddening and stellar crowding along each cluster's line--of--sight
reduced the member--to--target ratio to 8$\%$ for NGC 6342 and 37$\%$ for 
NGC 6366.  The cluster members provided average radial velocities of 
$+$112.5 km s$^{\rm -1}$ ($\sigma$ = 8.6 km s$^{\rm -1}$) and 
--122.3 km s$^{\rm -1}$ ($\sigma$ = 1.5 km s$^{\rm -1}$) for NGC 6342 and NGC 
6366, respectively.

From the sub--sample of confirmed cluster members, we were able to measure 
chemical abundances of O, Na, Mg, Si, Ca, Cr, Fe, Ni, and La for four stars in 
NGC 6342 and 13 stars in NGC 6366 via equivalent width measurements and 
spectrum synthesis fitting.  We find both clusters to have nearly identical
metallicities with NGC 6342 having $\langle$[Fe/H]$\rangle$ = --0.53 
($\sigma$ = 0.11) and NGC 6366 having 
$\langle$[Fe/H]$\rangle$ = --0.55 ($\sigma$ = 0.09).  Neither cluster shows 
significant evidence favoring a metallicity spread.  Both clusters exhibit 
very similar average [X/Fe] ratios and star--to--star abundance variations, 
but O and Na are likely the only two elements that exhibit
significant star--to--star scatter.  NGC 6366 shows evidence of only a 
moderately extended O--Na anti--correlation, but more data are needed for 
NGC 6342 to determine if this cluster also follows the same light element 
pattern.  The lack of additional abundance correlations in NGC 6366 
(e.g., Mg--Si correlation) indicates that the mechanism responsible for the 
O--Na anti--correlation did not reach temperatures high enough to significantly
deplete Mg nor produce Si.

Although [O/Fe] is significantly enhanced ([O/Fe] $>$ $+$0.50) for most stars 
in our sample, the heavier $\alpha$--elements have a more modest enhancement
of $\langle$[$\alpha$/Fe]$\rangle$ = $+$0.31 ($\sigma$ = 0.06).  The 
Fe--peak elements Cr and Ni mostly track Fe, but there is some weak evidence
that NGC 6366 may be slightly more enhanced in [Cr/Fe] and [Ni/Fe] than
NGC 6342.  Interestingly, NGC 6342 and NGC 6366 are both moderately enhanced
in La with $\langle$[La/Fe]$\rangle$ $\sim$ $+$0.20, which likely indicates 
some enrichment via the main s--process.  When the abundance patterns of 
NGC 6342 and NGC 6366 are compared with those of other similar metal--rich 
([Fe/H] $>$ --1) bulge clusters, we find that most metal--rich clusters share 
a common composition pattern.  However, we find some evidence favoring 
significant cluster--to--cluster variations in [La/Fe], which could 
be an indication that inner Galaxy globular cluster formation was a more 
heterogeneous process than is reflected in the $\alpha$--element chemistry.

A further comparison between metal--rich bulge globular clusters and bulge 
field stars with [Fe/H] $<$ --0.4 indicates that both populations exhibit 
nearly identical [$\alpha$/Fe], [Cr/Fe], and [Ni/Fe] abundance trends.  
However, the clusters are distinguished from the field stars by exhibiting 
larger [O/Fe] and [Na/Fe] dispersions, and also by their enhanced [La/Fe] 
abundances (for some clusters).  At [Fe/H] $>$ --0.4, the most metal--rich 
globular clusters may be further distinguished from the bulge field stars by 
remaining enhanced in [$\alpha$/Fe] up to at least solar metallicity.

\acknowledgements

This paper uses data products produced by the OIR Telescope Data Center, 
supported by the Smithsonian Astrophysical Observatory.  This research has 
made use of NASA's Astrophysics Data System Bibliographic Services.  This 
publication has made use of data products from the Two Micron All Sky Survey, 
which is a joint project of the University of Massachusetts and the Infrared 
Processing and Analysis Center/California Institute of Technology, funded by 
the National Aeronautics and Space Administration and the National Science 
Foundation.  The Second Palomar Observatory Sky Survey (POSS-II) was made by 
the California Institute of Technology with funds from the National Science 
Foundation, the National Geographic Society, the Sloan Foundation, the Samuel 
Oschin Foundation, and the Eastman Kodak Corporation.  C.I.J. gratefully 
acknowledges support from the Clay Fellowship, administered by the Smithsonian 
Astrophysical Observatory.  R.M.R. acknowledges support from the National 
Science Foundation (AST--1413755 and AST--1412673).  C.A.P. gratefully 
acknowledges support from the Daniel Kirkwood Research Fund at Indiana 
University and from the National Science Foundation (AST--1412673).

\clearpage

\clearpage

\clearpage

\clearpage

\clearpage

\clearpage

\clearpage
\tablenum{1}
\tablecolumns{8}
\tablewidth{0pt}

\begin{deluxetable}{cccccccc}
\tabletypesize{\scriptsize}
\tablecaption{NGC 6342 Coordinates, Photometry, and Velocities}
\tablehead{
\colhead{Star Name}     &
\colhead{RA (J2000)}      &
\colhead{DEC (J2000)}      &
\colhead{J}      &
\colhead{H}      &
\colhead{K$_{\rm S}$}      &
\colhead{RV$_{\rm helio.}$}      &
\colhead{RV Error}      \\
\colhead{(2MASS)}      &
\colhead{(Degrees)}      &
\colhead{(Degrees)}      &
\colhead{(mag.)}      &
\colhead{(mag.)}      &
\colhead{(mag.)}      &
\colhead{(km s$^{\rm -1}$)}      &
\colhead{(km s$^{\rm -1}$)}
}

\startdata
\hline
\multicolumn{8}{c}{Hectochelle Probable Members}       \\
\hline
17194058$-$1937038	&	259.919111	&	$-$19.617743	&	13.531	&	12.862	&	12.644	&	$+$102.06	&	0.30	\\
17195030$-$1927431	&	259.959610	&	$-$19.461985	&	13.234	&	12.509	&	12.320	&	$+$121.84	&	0.26	\\
17202716$-$1932143	&	260.113200	&	$-$19.537306	&	13.800	&	13.046	&	12.825	&	$+$110.83	&	0.49	\\
17205345$-$1957303	&	260.222729	&	$-$19.958424	&	12.477	&	11.643	&	11.493	&	$+$95.49	&	0.22	\\
17210009$-$1935354	&	260.250404	&	$-$19.593189	&	12.517	&	11.700	&	11.419	&	$+$121.35	&	0.24	\\
17210680$-$1935191	&	260.278335	&	$-$19.588644	&	13.071	&	12.292	&	12.093	&	$+$112.73	&	0.31	\\
17210888$-$1935147	&	260.287009	&	$-$19.587433	&	13.678	&	13.014	&	12.861	&	$+$117.16	&	0.43	\\
17211070$-$1935147	&	260.294586	&	$-$19.587420	&	13.545	&	12.936	&	12.649	&	$+$119.93	&	0.53	\\
17211222$-$1935291	&	260.300920	&	$-$19.591421	&	13.203	&	12.510	&	12.213	&	$+$115.39	&	0.32	\\
17214926$-$1947318	&	260.455264	&	$-$19.792168	&	13.556	&	12.856	&	12.728	&	$+$102.01	&	0.43	\\
17220959$-$1919193	&	260.539973	&	$-$19.322044	&	13.885	&	13.066	&	12.807	&	$+$99.59	&	1.06	\\
17221321$-$1934325	&	260.555060	&	$-$19.575714	&	12.969	&	12.175	&	11.990	&	$+$120.55	&	0.27	\\
17223024$-$1957315	&	260.626001	&	$-$19.958773	&	12.932	&	12.347	&	12.228	&	$+$105.08	&	0.32	\\
\hline
\multicolumn{8}{c}{Hectochelle Probable Non--Members}       \\
\hline
17192669$-$1938549	&	259.861219	&	$-$19.648594	&	13.622	&	12.915	&	12.721	&	$-$67.94	&	0.33	\\
17192877$-$1940358	&	259.869892	&	$-$19.676620	&	13.553	&	12.978	&	12.723	&	$-$15.17	&	0.27	\\
17193913$-$1950376	&	259.913066	&	$-$19.843796	&	12.223	&	11.535	&	11.419	&	$-$173.25	&	0.33	\\
17193986$-$1947439	&	259.916084	&	$-$19.795542	&	12.277	&	11.526	&	11.250	&	$-$44.77	&	0.29	\\
17194015$-$1941590	&	259.917329	&	$-$19.699745	&	12.600	&	11.804	&	11.568	&	$-$2.47	&	0.28	\\
17194441$-$1928060	&	259.935070	&	$-$19.468357	&	13.593	&	12.913	&	12.626	&	$-$45.04	&	0.32	\\
17194455$-$1935278	&	259.935628	&	$-$19.591057	&	13.712	&	12.995	&	12.782	&	$+$51.12	&	0.26	\\
17194631$-$1939438	&	259.942995	&	$-$19.662176	&	13.493	&	12.830	&	12.651	&	$+$60.64	&	0.30	\\
17194709$-$1953597	&	259.946240	&	$-$19.899942	&	13.423	&	12.791	&	12.561	&	$+$15.20	&	0.31	\\
17194833$-$1923503	&	259.951416	&	$-$19.397331	&	12.826	&	12.054	&	11.873	&	$+$14.80	&	0.20	\\
17194834$-$1948062	&	259.951444	&	$-$19.801746	&	13.289	&	12.612	&	12.378	&	$+$11.26	&	0.35	\\
17195217$-$1927124	&	259.967399	&	$-$19.453459	&	13.510	&	12.770	&	12.637	&	$+$64.17	&	0.27	\\
17195386$-$1932185	&	259.974422	&	$-$19.538490	&	13.636	&	12.966	&	12.797	&	$-$2.61	&	0.31	\\
17195710$-$1958071	&	259.987920	&	$-$19.968639	&	12.466	&	11.663	&	11.488	&	$+$92.36	&	0.23	\\
17195849$-$1950235	&	259.993709	&	$-$19.839867	&	13.558	&	12.895	&	12.734	&	$+$34.86	&	0.23	\\
17200094$-$1947559	&	260.003947	&	$-$19.798882	&	12.936	&	12.319	&	12.161	&	$-$0.02	&	0.25	\\
17200119$-$1927409	&	260.004979	&	$-$19.461382	&	13.731	&	12.988	&	12.858	&	$-$222.21	&	0.65	\\
17200325$-$1959267	&	260.013568	&	$-$19.990774	&	12.636	&	12.013	&	11.857	&	$-$23.40	&	0.21	\\
17200383$-$1949125	&	260.015992	&	$-$19.820152	&	12.602	&	11.778	&	11.550	&	$-$18.30	&	0.27	\\
17200391$-$1939333	&	260.016309	&	$-$19.659252	&	13.204	&	12.527	&	12.312	&	$-$77.76	&	0.29	\\
17200465$-$1925073	&	260.019413	&	$-$19.418722	&	13.608	&	12.950	&	12.735	&	$+$9.54	&	0.36	\\
17200541$-$1941188	&	260.022577	&	$-$19.688566	&	13.678	&	12.986	&	12.831	&	$-$67.42	&	0.43	\\
17200601$-$1944228	&	260.025057	&	$-$19.739676	&	13.535	&	12.905	&	12.682	&	$-$25.42	&	0.25	\\
17200681$-$1932100	&	260.028386	&	$-$19.536121	&	13.315	&	12.535	&	12.336	&	$+$89.66	&	0.34	\\
17200762$-$1942402	&	260.031790	&	$-$19.711184	&	13.620	&	12.922	&	12.689	&	$-$7.22	&	0.32	\\
17201041$-$1937471	&	260.043377	&	$-$19.629772	&	13.526	&	12.847	&	12.615	&	$-$1.74	&	0.31	\\
17201186$-$1938344	&	260.049441	&	$-$19.642895	&	13.674	&	12.931	&	12.746	&	$+$29.12	&	0.36	\\
17201294$-$1920323	&	260.053921	&	$-$19.342319	&	13.678	&	12.986	&	12.741	&	$+$30.09	&	0.32	\\
17201342$-$1954417	&	260.055927	&	$-$19.911585	&	12.274	&	11.534	&	11.258	&	$-$22.70	&	0.23	\\
17201592$-$1944023	&	260.066363	&	$-$19.733978	&	12.495	&	11.649	&	11.356	&	$-$14.71	&	0.38	\\
17201656$-$2001560	&	260.069029	&	$-$20.032248	&	13.307	&	12.555	&	12.276	&	$-$15.65	&	0.31	\\
17201818$-$1958002	&	260.075772	&	$-$19.966728	&	13.417	&	12.709	&	12.458	&	$+$30.73	&	0.30	\\
17202381$-$1940495	&	260.099231	&	$-$19.680441	&	12.563	&	11.775	&	11.549	&	$-$24.78	&	0.21	\\
17202393$-$1950597	&	260.099730	&	$-$19.849943	&	13.519	&	12.827	&	12.655	&	$+$59.11	&	0.26	\\
17202536$-$1934185	&	260.105708	&	$-$19.571814	&	12.401	&	11.590	&	11.404	&	$-$13.20	&	0.22	\\
17202587$-$1959192	&	260.107800	&	$-$19.988670	&	12.554	&	11.773	&	11.588	&	$-$186.46	&	0.25	\\
17202650$-$1945302	&	260.110432	&	$-$19.758400	&	13.090	&	12.428	&	12.166	&	$-$93.73	&	0.27	\\
17202717$-$1954207	&	260.113223	&	$-$19.905752	&	13.318	&	12.641	&	12.410	&	$-$15.98	&	0.32	\\
17202836$-$1947377	&	260.118180	&	$-$19.793812	&	13.596	&	12.890	&	12.677	&	$-$0.33	&	0.24	\\
17202890$-$1928013	&	260.120424	&	$-$19.467030	&	13.676	&	12.995	&	12.720	&	$+$6.67	&	0.43	\\
17202903$-$1952412	&	260.120978	&	$-$19.878136	&	13.619	&	12.919	&	12.725	&	$-$197.44	&	0.34	\\
17203084$-$1920274	&	260.128525	&	$-$19.340967	&	13.798	&	13.014	&	12.872	&	$-$16.72	&	0.40	\\
17203202$-$1952211	&	260.133456	&	$-$19.872538	&	13.828	&	13.074	&	12.890	&	$+$31.95	&	0.24	\\
17203287$-$1958005	&	260.136992	&	$-$19.966810	&	12.383	&	11.560	&	11.368	&	$+$67.02	&	0.36	\\
17203363$-$1925030	&	260.140136	&	$-$19.417503	&	13.742	&	13.009	&	12.789	&	$-$2.57	&	0.31	\\
17203422$-$1926568	&	260.142600	&	$-$19.449112	&	13.024	&	12.239	&	12.062	&	$-$106.07	&	0.36	\\
17203468$-$1916379	&	260.144507	&	$-$19.277214	&	13.667	&	13.011	&	12.747	&	$-$25.50	&	0.35	\\
17203485$-$1943309	&	260.145247	&	$-$19.725267	&	13.698	&	13.015	&	12.832	&	$+$24.77	&	0.32	\\
17203825$-$1934067	&	260.159379	&	$-$19.568529	&	13.523	&	12.823	&	12.642	&	$+$70.73	&	0.32	\\
17203875$-$1945043	&	260.161478	&	$-$19.751204	&	12.849	&	12.139	&	11.891	&	$-$42.01	&	0.28	\\
17203889$-$1944226	&	260.162082	&	$-$19.739637	&	12.771	&	11.965	&	11.752	&	$+$33.21	&	0.22	\\
17203957$-$1939576	&	260.164897	&	$-$19.666012	&	13.223	&	12.470	&	12.251	&	$+$25.22	&	0.28	\\
17204035$-$1919530	&	260.168128	&	$-$19.331398	&	12.408	&	11.442	&	11.202	&	$-$90.75	&	0.28	\\
17204113$-$1916533	&	260.171412	&	$-$19.281488	&	13.262	&	12.533	&	12.316	&	$-$48.48	&	0.27	\\
17204129$-$1936548	&	260.172077	&	$-$19.615242	&	12.489	&	11.568	&	11.301	&	$+$13.79	&	0.26	\\
17204195$-$1938438	&	260.174820	&	$-$19.645512	&	12.844	&	12.046	&	11.882	&	$+$74.29	&	0.24	\\
17204237$-$1924343	&	260.176579	&	$-$19.409550	&	13.852	&	13.117	&	12.864	&	$-$57.67	&	0.37	\\
17204339$-$1931097	&	260.180814	&	$-$19.519388	&	13.324	&	12.505	&	12.318	&	$+$2.89	&	0.33	\\
17204360$-$1944436	&	260.181667	&	$-$19.745449	&	13.547	&	12.896	&	12.689	&	$-$98.16	&	0.33	\\
17204374$-$1937366	&	260.182267	&	$-$19.626841	&	13.648	&	12.878	&	12.670	&	$-$100.61	&	0.30	\\
17204438$-$1927492	&	260.184931	&	$-$19.463688	&	13.544	&	12.833	&	12.670	&	$+$33.23	&	0.58	\\
17204731$-$1937150	&	260.197155	&	$-$19.620840	&	13.405	&	12.697	&	12.495	&	$-$8.60	&	0.34	\\
17204737$-$1953483	&	260.197402	&	$-$19.896757	&	13.229	&	12.475	&	12.331	&	$+$19.21	&	0.25	\\
17204822$-$1936487	&	260.200942	&	$-$19.613537	&	13.568	&	12.855	&	12.667	&	$+$46.87	&	0.30	\\
17204874$-$1941343	&	260.203120	&	$-$19.692863	&	13.279	&	12.488	&	12.281	&	$+$2.95	&	0.24	\\
17205014$-$1950474	&	260.208926	&	$-$19.846519	&	13.262	&	12.574	&	12.325	&	$-$26.47	&	0.35	\\
17205047$-$1940358	&	260.210316	&	$-$19.676624	&	13.493	&	12.808	&	12.583	&	$+$59.80	&	0.36	\\
17205126$-$1931479	&	260.213622	&	$-$19.529985	&	13.376	&	12.585	&	12.413	&	$+$278.60	&	0.39	\\
17205158$-$1948535	&	260.214935	&	$-$19.814873	&	13.332	&	12.606	&	12.456	&	$+$28.55	&	0.23	\\
17205172$-$1956596	&	260.215538	&	$-$19.949915	&	13.687	&	13.005	&	12.827	&	$-$12.20	&	0.33	\\
17205198$-$1941546	&	260.216605	&	$-$19.698515	&	13.727	&	13.014	&	12.835	&	$+$35.55	&	0.31	\\
17205293$-$1945365	&	260.220544	&	$-$19.760143	&	13.592	&	12.909	&	12.771	&	$-$20.29	&	0.31	\\
17205317$-$1945010	&	260.221564	&	$-$19.750296	&	13.429	&	12.752	&	12.555	&	$-$51.54	&	0.32	\\
17205330$-$1959442	&	260.222098	&	$-$19.995638	&	13.249	&	12.581	&	12.483	&	$-$50.19	&	0.26	\\
17205340$-$1936577	&	260.222534	&	$-$19.616049	&	13.398	&	12.708	&	12.510	&	$-$6.84	&	0.35	\\
17205462$-$1942107	&	260.227589	&	$-$19.702976	&	12.633	&	11.822	&	11.634	&	$+$26.84	&	0.35	\\
17205480$-$1943031	&	260.228360	&	$-$19.717537	&	13.358	&	12.773	&	12.585	&	$+$45.18	&	0.27	\\
17205483$-$1949197	&	260.228493	&	$-$19.822151	&	13.002	&	12.263	&	12.062	&	$-$53.58	&	0.31	\\
17205518$-$1931224	&	260.229952	&	$-$19.522907	&	13.643	&	12.869	&	12.726	&	$+$18.38	&	0.38	\\
17205620$-$1936570	&	260.234206	&	$-$19.615856	&	13.013	&	12.233	&	12.109	&	$-$25.43	&	0.24	\\
17205642$-$1925420	&	260.235108	&	$-$19.428358	&	13.706	&	13.263	&	12.877	&	$+$68.23	&	0.51	\\
17205688$-$1944533	&	260.237022	&	$-$19.748152	&	13.207	&	12.324	&	12.176	&	$+$32.45	&	1.67	\\
17205703$-$2004175	&	260.237643	&	$-$20.071552	&	13.465	&	12.830	&	12.704	&	$+$23.79	&	0.30	\\
17205715$-$1936317	&	260.238142	&	$-$19.608824	&	13.559	&	12.853	&	12.695	&	$-$22.97	&	0.28	\\
17205764$-$1931174	&	260.240187	&	$-$19.521500	&	13.600	&	12.819	&	12.566	&	$-$30.23	&	0.40	\\
17205865$-$1939355	&	260.244384	&	$-$19.659876	&	13.680	&	13.061	&	12.773	&	$-$15.14	&	0.56	\\
17205971$-$1936344	&	260.248796	&	$-$19.609556	&	13.288	&	12.540	&	12.376	&	$+$8.02	&	0.25	\\
17210143$-$1950243	&	260.255999	&	$-$19.840107	&	13.740	&	13.127	&	12.888	&	$+$35.36	&	0.28	\\
17210147$-$1957562	&	260.256152	&	$-$19.965622	&	13.511	&	12.930	&	12.748	&	$-$23.86	&	0.33	\\
17210222$-$1918201	&	260.259271	&	$-$19.305592	&	12.417	&	11.572	&	11.309	&	$+$24.19	&	0.31	\\
17210297$-$1951122	&	260.262381	&	$-$19.853411	&	13.753	&	13.101	&	12.844	&	$-$39.86	&	0.25	\\
17210303$-$1943136	&	260.262632	&	$-$19.720457	&	13.600	&	12.900	&	12.659	&	$-$0.92	&	0.36	\\
17210413$-$1920441	&	260.267222	&	$-$19.345585	&	13.444	&	12.677	&	12.462	&	$-$30.48	&	0.35	\\
17210455$-$1958310	&	260.268982	&	$-$19.975292	&	13.561	&	12.868	&	12.621	&	$-$28.84	&	0.33	\\
17210512$-$1940473	&	260.271356	&	$-$19.679811	&	12.310	&	11.448	&	11.216	&	$+$8.13	&	0.30	\\
17210526$-$1940225	&	260.271950	&	$-$19.672943	&	13.089	&	12.380	&	12.182	&	$-$90.80	&	0.22	\\
17210565$-$1943381	&	260.273581	&	$-$19.727251	&	13.614	&	12.876	&	12.674	&	$-$36.29	&	0.32	\\
17210752$-$1950133	&	260.281366	&	$-$19.837049	&	13.755	&	13.043	&	12.854	&	$+$8.49	&	0.28	\\
17210798$-$1944480	&	260.283250	&	$-$19.746670	&	12.733	&	12.021	&	11.789	&	$-$37.42	&	0.23	\\
17210937$-$1920314	&	260.289060	&	$-$19.342075	&	13.801	&	13.072	&	12.814	&	$-$106.74	&	0.38	\\
17210937$-$1952010	&	260.289051	&	$-$19.866949	&	12.709	&	11.926	&	11.669	&	$-$6.26	&	0.27	\\
17211069$-$1958044	&	260.294542	&	$-$19.967916	&	12.800	&	12.021	&	11.799	&	$+$12.96	&	0.25	\\
17211091$-$1920313	&	260.295466	&	$-$19.342045	&	12.612	&	11.798	&	11.543	&	$-$60.21	&	0.27	\\
17211110$-$1952125	&	260.296252	&	$-$19.870165	&	13.108	&	12.342	&	12.095	&	$+$27.07	&	0.25	\\
17211237$-$1950107	&	260.301579	&	$-$19.836317	&	13.715	&	13.071	&	12.881	&	$+$161.90	&	0.35	\\
17211260$-$1940190	&	260.302505	&	$-$19.671967	&	12.970	&	12.265	&	12.054	&	$+$70.92	&	0.22	\\
17211346$-$1937077	&	260.306116	&	$-$19.618828	&	13.570	&	12.882	&	12.732	&	$-$53.72	&	0.36	\\
17211365$-$1948387	&	260.306893	&	$-$19.810772	&	13.589	&	12.876	&	12.674	&	$+$7.63	&	0.35	\\
17211399$-$1945176	&	260.308319	&	$-$19.754894	&	13.586	&	12.926	&	12.740	&	$-$23.56	&	0.31	\\
17211457$-$1935059	&	260.310717	&	$-$19.584993	&	13.709	&	12.937	&	12.764	&	$-$87.90	&	0.26	\\
17211567$-$1925399	&	260.315300	&	$-$19.427755	&	13.546	&	12.835	&	12.644	&	$-$5.09	&	0.37	\\
17211579$-$1920207	&	260.315820	&	$-$19.339090	&	13.449	&	12.756	&	12.595	&	$+$28.27	&	0.45	\\
17211604$-$1958507	&	260.316866	&	$-$19.980768	&	13.470	&	12.684	&	12.525	&	$-$1.61	&	0.28	\\
17211625$-$1940345	&	260.317715	&	$-$19.676260	&	12.550	&	11.729	&	11.549	&	$-$116.56	&	0.22	\\
17211742$-$1945086	&	260.322603	&	$-$19.752401	&	13.396	&	12.759	&	12.524	&	$-$85.73	&	0.32	\\
17211786$-$1942484	&	260.324441	&	$-$19.713463	&	13.584	&	12.856	&	12.681	&	$+$64.33	&	0.26	\\
17211791$-$1944255	&	260.324638	&	$-$19.740423	&	13.696	&	12.956	&	12.724	&	$-$23.85	&	0.31	\\
17211818$-$1942038	&	260.325764	&	$-$19.701065	&	13.441	&	12.728	&	12.581	&	$-$27.01	&	0.29	\\
17211853$-$1916328	&	260.327237	&	$-$19.275803	&	12.782	&	11.999	&	11.783	&	$-$60.06	&	0.42	\\
17211928$-$1933077	&	260.330352	&	$-$19.552147	&	13.296	&	12.602	&	12.384	&	$+$15.96	&	0.30	\\
17212066$-$1936171	&	260.336095	&	$-$19.604750	&	13.778	&	13.113	&	12.840	&	$+$253.12	&	2.20	\\
17212068$-$1939312	&	260.336184	&	$-$19.658682	&	12.987	&	12.251	&	12.049	&	$-$5.44	&	0.24	\\
17212110$-$1931128	&	260.337920	&	$-$19.520243	&	13.736	&	13.016	&	12.810	&	$-$42.87	&	0.35	\\
17212113$-$1925169	&	260.338052	&	$-$19.421383	&	13.553	&	12.787	&	12.600	&	$+$57.42	&	0.41	\\
17212139$-$1934169\tablenotemark{a}	&	260.339162	&	$-$19.571383	&	12.314	&	11.522	&	11.307	&	$+$2.36	&	0.21	\\
17212205$-$1949292	&	260.341878	&	$-$19.824793	&	13.540	&	12.813	&	12.613	&	$+$1.58	&	0.31	\\
17212277$-$1918506	&	260.344898	&	$-$19.314058	&	13.716	&	12.975	&	12.804	&	$-$147.47	&	0.57	\\
17212453$-$1957342	&	260.352217	&	$-$19.959501	&	13.468	&	12.784	&	12.542	&	$-$63.53	&	0.32	\\
17212549$-$1933147	&	260.356213	&	$-$19.554100	&	12.897	&	12.120	&	11.886	&	$-$55.29	&	0.25	\\
17212577$-$1946584	&	260.357394	&	$-$19.782894	&	13.076	&	12.279	&	11.997	&	$+$7.83	&	0.39	\\
17212641$-$1954475	&	260.360066	&	$-$19.913216	&	13.596	&	12.924	&	12.706	&	$-$97.17	&	0.31	\\
17212673$-$1937056	&	260.361379	&	$-$19.618248	&	13.671	&	13.009	&	12.825	&	$-$21.95	&	0.27	\\
17212695$-$1951540	&	260.362309	&	$-$19.865021	&	13.214	&	12.547	&	12.323	&	$-$65.55	&	0.29	\\
17212727$-$1919391	&	260.363638	&	$-$19.327541	&	13.484	&	12.698	&	12.470	&	$-$71.42	&	0.34	\\
17212772$-$1942170	&	260.365508	&	$-$19.704741	&	13.482	&	12.749	&	12.540	&	$-$17.89	&	0.60	\\
17212896$-$1944170	&	260.370675	&	$-$19.738064	&	13.576	&	12.818	&	12.669	&	$-$21.93	&	0.22	\\
17212920$-$1922228	&	260.371674	&	$-$19.373022	&	12.852	&	11.980	&	11.750	&	$-$17.40	&	0.25	\\
17212931$-$1950420	&	260.372135	&	$-$19.845011	&	13.842	&	13.128	&	12.891	&	$-$20.52	&	0.32	\\
17213073$-$1945299	&	260.378043	&	$-$19.758329	&	13.331	&	12.608	&	12.327	&	$-$38.51	&	0.33	\\
17213179$-$1943154	&	260.382474	&	$-$19.720964	&	12.952	&	12.112	&	11.851	&	$-$15.17	&	0.28	\\
17213229$-$1921190	&	260.384576	&	$-$19.355305	&	13.447	&	12.679	&	12.407	&	$-$35.55	&	0.36	\\
17213359$-$1940422\tablenotemark{a}	&	260.389966	&	$-$19.678402	&	12.250	&	11.498	&	11.228	&	$+$52.92	&	0.33	\\
17213428$-$1932347	&	260.392844	&	$-$19.542984	&	13.081	&	12.280	&	12.065	&	$+$147.23	&	0.34	\\
17213428$-$1949324	&	260.392865	&	$-$19.825682	&	13.519	&	12.841	&	12.672	&	$+$20.98	&	0.34	\\
17213527$-$1937276	&	260.396978	&	$-$19.624346	&	13.410	&	12.704	&	12.505	&	$+$82.25	&	0.26	\\
17213535$-$1933175	&	260.397293	&	$-$19.554869	&	13.678	&	13.030	&	12.840	&	$-$87.71	&	0.35	\\
17213588$-$1938294	&	260.399500	&	$-$19.641527	&	13.710	&	13.066	&	12.878	&	$+$57.45	&	0.30	\\
17213620$-$1933593	&	260.400870	&	$-$19.566496	&	13.150	&	12.443	&	12.176	&	$+$56.97	&	0.38	\\
17213681$-$1941537	&	260.403383	&	$-$19.698273	&	13.600	&	12.921	&	12.658	&	$-$3.25	&	0.38	\\
17213687$-$1959067	&	260.403634	&	$-$19.985216	&	12.338	&	11.555	&	11.292	&	$-$6.78	&	0.22	\\
17213801$-$1941038	&	260.408391	&	$-$19.684408	&	13.508	&	12.749	&	12.501	&	$-$22.99	&	0.39	\\
17213875$-$1949446	&	260.411475	&	$-$19.829069	&	13.706	&	13.099	&	12.853	&	$+$28.54	&	0.29	\\
17213914$-$2003403	&	260.413099	&	$-$20.061220	&	13.078	&	12.406	&	12.168	&	$+$40.69	&	0.32	\\
17214034$-$1927316	&	260.418101	&	$-$19.458784	&	13.358	&	12.583	&	12.317	&	$-$84.65	&	0.43	\\
17214107$-$1929095	&	260.421147	&	$-$19.485977	&	13.617	&	12.843	&	12.632	&	$-$7.10	&	0.34	\\
17214286$-$1958157	&	260.428624	&	$-$19.971031	&	13.766	&	13.140	&	12.888	&	$+$26.34	&	0.36	\\
17214319$-$1942560	&	260.429993	&	$-$19.715557	&	13.178	&	12.397	&	12.187	&	$+$53.38	&	0.27	\\
17214324$-$1938210	&	260.430182	&	$-$19.639181	&	13.564	&	12.935	&	12.728	&	$-$43.56	&	0.33	\\
17214327$-$1951540	&	260.430295	&	$-$19.865026	&	13.673	&	12.981	&	12.743	&	$-$15.59	&	0.34	\\
17214398$-$1943354	&	260.433275	&	$-$19.726515	&	12.901	&	12.156	&	11.891	&	$-$0.56	&	0.26	\\
17214502$-$1934383	&	260.437585	&	$-$19.577307	&	13.291	&	12.503	&	12.288	&	$+$9.14	&	0.26	\\
17214584$-$1955322	&	260.441008	&	$-$19.925638	&	13.271	&	12.466	&	12.242	&	$-$26.24	&	0.31	\\
17214647$-$1941436	&	260.443641	&	$-$19.695459	&	13.640	&	12.922	&	12.753	&	$+$42.66	&	0.30	\\
17214698$-$1933032	&	260.445753	&	$-$19.550915	&	13.653	&	13.006	&	12.830	&	$+$46.55	&	0.32	\\
17214700$-$1939214	&	260.445845	&	$-$19.655970	&	13.472	&	12.787	&	12.564	&	$-$34.49	&	0.34	\\
17214767$-$1927419	&	260.448636	&	$-$19.461641	&	13.253	&	12.529	&	12.273	&	$+$15.03	&	0.34	\\
17214791$-$1929004	&	260.449653	&	$-$19.483463	&	13.015	&	12.352	&	12.090	&	$+$23.06	&	0.31	\\
17215020$-$1948473	&	260.459183	&	$-$19.813160	&	12.449	&	11.681	&	11.460	&	$+$195.38	&	0.32	\\
17215025$-$1930586	&	260.459411	&	$-$19.516294	&	12.338	&	11.525	&	11.286	&	$-$140.92	&	0.23	\\
17215274$-$1946097	&	260.469781	&	$-$19.769363	&	12.928	&	12.138	&	11.944	&	$+$166.62	&	0.28	\\
17215378$-$1959302	&	260.474088	&	$-$19.991728	&	13.603	&	12.946	&	12.783	&	$+$54.49	&	0.30	\\
17215554$-$1950411	&	260.481430	&	$-$19.844772	&	13.190	&	12.518	&	12.313	&	$+$30.49	&	0.31	\\
17215764$-$1955389	&	260.490197	&	$-$19.927483	&	13.698	&	12.999	&	12.787	&	$+$8.15	&	0.30	\\
17215768$-$1921148	&	260.490369	&	$-$19.354113	&	13.083	&	12.307	&	12.106	&	$-$16.66	&	0.39	\\
17220087$-$1943226	&	260.503649	&	$-$19.722956	&	13.751	&	13.036	&	12.804	&	$+$87.31	&	0.30	\\
17220161$-$1942348	&	260.506713	&	$-$19.709694	&	12.722	&	11.960	&	11.715	&	$-$5.12	&	0.25	\\
17220360$-$1937208	&	260.515012	&	$-$19.622456	&	13.658	&	13.027	&	12.799	&	$-$30.85	&	0.30	\\
17220422$-$1953478	&	260.517594	&	$-$19.896614	&	13.440	&	12.744	&	12.520	&	$+$8.65	&	0.33	\\
17220497$-$1939260	&	260.520711	&	$-$19.657248	&	13.690	&	12.961	&	12.746	&	$+$2.38	&	1.20	\\
17220535$-$1949517	&	260.522326	&	$-$19.831038	&	13.465	&	12.715	&	12.525	&	$-$10.79	&	0.31	\\
17220536$-$1933156	&	260.522334	&	$-$19.554342	&	13.467	&	12.747	&	12.559	&	$-$24.87	&	0.28	\\
17220623$-$1925040	&	260.525982	&	$-$19.417786	&	12.809	&	12.026	&	11.782	&	$-$80.17	&	0.32	\\
17220941$-$1937289	&	260.539223	&	$-$19.624699	&	13.530	&	12.804	&	12.540	&	$-$25.75	&	0.39	\\
17220978$-$1940206	&	260.540764	&	$-$19.672413	&	13.066	&	12.355	&	12.136	&	$-$71.59	&	0.33	\\
17221115$-$1931581\tablenotemark{a}	&	260.546495	&	$-$19.532812	&	12.583	&	11.710	&	11.475	&	$+$27.44	&	0.21	\\
17221579$-$1952348	&	260.565816	&	$-$19.876337	&	13.626	&	12.911	&	12.691	&	$-$28.71	&	0.32	\\
17221708$-$1948102	&	260.571187	&	$-$19.802839	&	12.425	&	11.633	&	11.374	&	$+$55.33	&	0.28	\\
17221779$-$1935538	&	260.574161	&	$-$19.598288	&	12.854	&	12.065	&	11.815	&	$+$155.68	&	0.27	\\
17222021$-$1951100	&	260.584248	&	$-$19.852785	&	13.791	&	13.129	&	12.845	&	$-$113.35	&	0.33	\\
17222304$-$1950369	&	260.596018	&	$-$19.843597	&	13.130	&	12.354	&	12.083	&	$-$91.29	&	0.26	\\
17222359$-$1943386	&	260.598299	&	$-$19.727390	&	13.739	&	13.107	&	12.840	&	$-$83.87	&	0.35	\\
17222641$-$1931500	&	260.610063	&	$-$19.530565	&	12.913	&	12.077	&	11.815	&	$+$47.15	&	0.44	\\
17222780$-$1945183	&	260.615868	&	$-$19.755096	&	13.515	&	12.860	&	12.642	&	$+$70.22	&	0.41	\\
17222890$-$1921019	&	260.620446	&	$-$19.350554	&	12.763	&	12.161	&	11.942	&	$+$25.50	&	0.37	\\
17223138$-$1942206	&	260.630788	&	$-$19.705727	&	13.283	&	12.564	&	12.379	&	$-$125.03	&	1.16	\\
17223206$-$1927114	&	260.633603	&	$-$19.453180	&	13.688	&	13.066	&	12.836	&	$-$75.33	&	0.63	\\
17223532$-$1941053	&	260.647189	&	$-$19.684830	&	13.666	&	12.955	&	12.758	&	$-$82.18	&	0.26	\\
17223662$-$1927567	&	260.652614	&	$-$19.465775	&	12.350	&	11.736	&	11.567	&	$-$84.16	&	0.38	\\
17224344$-$1934285	&	260.681014	&	$-$19.574606	&	13.282	&	12.498	&	12.234	&	$+$16.75	&	0.32	\\
17224348$-$1945514	&	260.681206	&	$-$19.764290	&	12.677	&	11.868	&	11.655	&	$-$39.80	&	0.23	\\
17224880$-$1941344	&	260.703367	&	$-$19.692913	&	13.748	&	13.050	&	12.837	&	$+$18.78	&	0.46	\\
17225118$-$1939432	&	260.713283	&	$-$19.662014	&	13.572	&	12.873	&	12.579	&	$-$71.51	&	0.32	\\
17225880$-$1930327	&	260.745003	&	$-$19.509094	&	13.758	&	13.040	&	12.811	&	$+$8.35	&	0.61	\\
\hline
\multicolumn{8}{c}{Hydra Probable Members}       \\
\hline
17203668$-$1939270	&	260.152851	&	$-$19.657518	&	11.798	&	10.868	&	10.674	&	$+$110.01	&	0.28	\\
17205974$-$1939237	&	260.248945	&	$-$19.656588	&	11.319	&	10.451	&	10.086	&	$+$98.28	&	0.71	\\
17210211$-$1932252	&	260.258823	&	$-$19.540339	&	12.396	&	11.542	&	11.257	&	$+$113.28	&	0.66	\\
17210615$-$1934073	&	260.275637	&	$-$19.568710	&	12.502	&	11.600	&	11.347	&	$+$117.27	&	0.43	\\
17210813$-$1933174	&	260.283905	&	$-$19.554838	&	12.311	&	11.399	&	11.157	&	$+$115.40	&	0.87	\\
17211015$-$1935227	&	260.292307	&	$-$19.589666	&	11.901	&	11.098	&	10.809	&	$+$126.56	&	0.29	\\
17211185$-$1934551	&	260.299401	&	$-$19.581997	&	11.915	&	10.991	&	10.744	&	$+$115.25	&	0.30	\\
17213696$-$1945045	&	260.404032	&	$-$19.751265	&	12.470	&	11.661	&	11.433	&	$+$116.28	&	0.30	\\
17220360$-$1933439	&	260.515039	&	$-$19.562201	&	12.426	&	11.616	&	11.335	&	$+$119.50	&	0.62	\\
\hline
\multicolumn{8}{c}{Hydra Probable Non--Members}       \\
\hline
17201119$-$1932228	&	260.046651	&	$-$19.539692	&	12.317	&	11.456	&	11.206	&	$+$32.13	&	0.40	\\
17201643$-$1934577	&	260.068491	&	$-$19.582697	&	11.072	&	10.178	&	9.843	&	$-$16.76	&	0.27	\\
17202313$-$1937485	&	260.096390	&	$-$19.630148	&	11.213	&	10.299	&	10.023	&	$+$27.19	&	0.48	\\
17202663$-$1929066	&	260.110969	&	$-$19.485168	&	12.405	&	11.512	&	11.243	&	$-$61.52	&	0.53	\\
17202927$-$1944416	&	260.121986	&	$-$19.744911	&	11.558	&	10.672	&	10.378	&	$+$31.90	&	0.52	\\
17203014$-$1930574	&	260.125610	&	$-$19.515949	&	11.461	&	10.564	&	10.223	&	$+$13.95	&	0.43	\\
17203482$-$1935256	&	260.145119	&	$-$19.590456	&	11.912	&	11.021	&	10.723	&	$-$9.88	&	0.50	\\
17203834$-$1925180	&	260.159767	&	$-$19.421684	&	12.090	&	11.152	&	10.885	&	$-$89.51	&	1.47	\\
17203848$-$1945061	&	260.160336	&	$-$19.751698	&	12.473	&	11.737	&	11.433	&	$+$55.49	&	0.73	\\
17204202$-$1940132	&	260.175116	&	$-$19.670353	&	12.616	&	11.744	&	11.476	&	$-$46.53	&	0.43	\\
17205272$-$1947140	&	260.219697	&	$-$19.787228	&	12.061	&	11.289	&	11.019	&	$+$40.73	&	0.50	\\
17210354$-$1920423	&	260.264787	&	$-$19.345095	&	12.431	&	11.563	&	11.331	&	$+$68.03	&	0.24	\\
17210404$-$1948513	&	260.266848	&	$-$19.814257	&	11.249	&	10.417	&	10.079	&	$-$47.61	&	1.06	\\
17210448$-$1931152	&	260.268691	&	$-$19.520906	&	12.331	&	11.673	&	11.307	&	$+$6.37	&	0.38	\\
17211406$-$1928370	&	260.308584	&	$-$19.476955	&	11.802	&	10.950	&	10.625	&	$+$24.99	&	0.37	\\
17211414$-$1936108	&	260.308928	&	$-$19.603008	&	12.380	&	11.535	&	11.260	&	$+$5.47	&	0.46	\\
17211816$-$1920425	&	260.325674	&	$-$19.345148	&	12.221	&	11.346	&	11.054	&	$+$22.97	&	0.46	\\
17211820$-$1931065	&	260.325854	&	$-$19.518476	&	12.280	&	11.478	&	11.254	&	$-$20.44	&	0.31	\\
17212036$-$1927053	&	260.334836	&	$-$19.451488	&	12.164	&	11.250	&	10.993	&	$+$40.02	&	0.48	\\
17212047$-$1947445	&	260.335316	&	$-$19.795708	&	11.709	&	10.886	&	10.621	&	$-$43.53	&	0.20	\\
17212084$-$1941512	&	260.336836	&	$-$19.697559	&	11.436	&	10.551	&	10.311	&	$+$70.53	&	0.41	\\
17212139$-$1934169\tablenotemark{a}	&	260.339162	&	$-$19.571383	&	12.314	&	11.522	&	11.307	&	$+$1.25	&	0.22	\\
17212508$-$1937461	&	260.354541	&	$-$19.629480	&	12.314	&	11.436	&	11.241	&	$+$22.87	&	0.28	\\
17212583$-$1947142	&	260.357646	&	$-$19.787296	&	12.191	&	11.366	&	11.103	&	$-$13.98	&	0.45	\\
17212689$-$1931313	&	260.362079	&	$-$19.525373	&	11.357	&	10.476	&	10.143	&	$-$23.77	&	0.83	\\
17212880$-$1944224	&	260.370018	&	$-$19.739576	&	12.610	&	11.679	&	11.483	&	$+$7.67	&	0.31	\\
17213067$-$1943533	&	260.377827	&	$-$19.731485	&	11.922	&	11.084	&	10.797	&	$-$41.75	&	0.31	\\
17213310$-$1932538	&	260.387926	&	$-$19.548296	&	11.556	&	10.725	&	10.434	&	$-$34.14	&	0.63	\\
17213359$-$1940422\tablenotemark{a}	&	260.389966	&	$-$19.678402	&	12.250	&	11.498	&	11.228	&	$+$46.82	&	0.35	\\
17213753$-$1936323	&	260.406375	&	$-$19.608999	&	11.741	&	10.937	&	10.654	&	$-$6.15	&	0.43	\\
17214524$-$1924304	&	260.438500	&	$-$19.408463	&	12.581	&	11.721	&	11.445	&	$-$155.32	&	0.30	\\
17214945$-$1944587	&	260.456078	&	$-$19.749660	&	11.054	&	10.165	&	9.855	&	$-$44.25	&	0.41	\\
17215025$-$1925417	&	260.459397	&	$-$19.428253	&	12.244	&	11.365	&	11.141	&	$+$82.48	&	0.32	\\
17215072$-$1927350	&	260.461360	&	$-$19.459749	&	12.345	&	11.520	&	11.254	&	$+$51.74	&	0.73	\\
17215536$-$1939570	&	260.480698	&	$-$19.665852	&	11.248	&	10.376	&	10.121	&	$+$90.28	&	0.24	\\
17215932$-$1926573	&	260.497205	&	$-$19.449261	&	12.064	&	11.217	&	10.890	&	$-$1.66	&	0.58	\\
17215984$-$1936324	&	260.499343	&	$-$19.609022	&	12.391	&	11.551	&	11.351	&	$+$68.20	&	0.26	\\
17220005$-$1930599	&	260.500227	&	$-$19.516642	&	11.006	&	10.039	&	9.750	&	$-$48.17	&	0.57	\\
17220327$-$1935141	&	260.513647	&	$-$19.587261	&	11.384	&	10.536	&	10.214	&	$-$32.14	&	0.59	\\
17220575$-$1927591	&	260.523999	&	$-$19.466438	&	12.360	&	11.566	&	11.347	&	$+$55.41	&	0.65	\\
17220656$-$1941191	&	260.527372	&	$-$19.688658	&	11.143	&	10.262	&	9.936	&	$+$75.73	&	0.65	\\
17221115$-$1931581\tablenotemark{a}	&	260.546495	&	$-$19.532812	&	12.583	&	11.710	&	11.475	&	$+$27.26	&	0.28	\\
\enddata

\tablenotetext{a}{This flag indicates that the star was observed with both
the Hectochelle and Hydra instruments.}

\end{deluxetable}

\clearpage
\tablenum{2}
\tablecolumns{8}
\tablewidth{0pt}

\begin{deluxetable}{cccccccc}
\tabletypesize{\scriptsize}
\tablecaption{NGC 6366 Coordinates, Photometry, and Velocities}
\tablehead{
\colhead{Star Name}     &
\colhead{RA (J2000)}      &
\colhead{DEC (J2000)}      &
\colhead{J}      &
\colhead{H}      &
\colhead{K$_{\rm S}$}      &
\colhead{RV$_{\rm helio.}$}      &
\colhead{RV Error}      \\
\colhead{(2MASS)}      &
\colhead{(Degrees)}      &
\colhead{(Degrees)}      &
\colhead{(mag.)}      &
\colhead{(mag.)}      &
\colhead{(mag.)}      &
\colhead{(km s$^{\rm -1}$)}      &
\colhead{(km s$^{\rm -1}$)}
}

\startdata
\hline
\multicolumn{8}{c}{Hydra Probable Members}       \\
\hline
17271061$-$0457415	&	261.794225	&	$-$4.961547	&	11.016	&	10.189	&	9.929	&	$-$121.91	&	0.21	\\
17272071$-$0505087	&	261.836299	&	$-$5.085757	&	11.288	&	10.482	&	10.278	&	$-$122.47	&	0.21	\\
17272367$-$0502272	&	261.848632	&	$-$5.040897	&	10.138	&	9.200	&	8.918	&	$-$121.24	&	0.22	\\
17273010$-$0504197	&	261.875443	&	$-$5.072152	&	10.860	&	9.994	&	9.747	&	$-$123.89	&	0.22	\\
17273285$-$0500304	&	261.886891	&	$-$5.008450	&	11.140	&	10.316	&	10.105	&	$-$119.99	&	0.22	\\
17274128$-$0505308	&	261.922036	&	$-$5.091896	&	11.311	&	10.506	&	10.255	&	$-$124.17	&	0.20	\\
17274221$-$0506173	&	261.925909	&	$-$5.104828	&	10.061	&	9.131	&	8.872	&	$-$121.33	&	0.22	\\
17274279$-$0504077	&	261.928333	&	$-$5.068811	&	10.930	&	10.041	&	9.829	&	$-$124.53	&	0.22	\\
17274452$-$0502371	&	261.935500	&	$-$5.043651	&	10.985	&	10.157	&	9.946	&	$-$122.38	&	0.21	\\
17274541$-$0504089	&	261.939246	&	$-$5.069143	&	11.121	&	10.283	&	10.052	&	$-$123.45	&	0.20	\\
17274724$-$0500362	&	261.946845	&	$-$5.010075	&	10.620	&	9.714	&	9.458	&	$-$123.22	&	0.21	\\
17274809$-$0507395	&	261.950399	&	$-$5.127661	&	11.183	&	10.296	&	10.093	&	$-$122.99	&	0.24	\\
17274954$-$0512022	&	261.956422	&	$-$5.200619	&	9.833	&	8.853	&	8.583	&	$-$123.45	&	0.24	\\
17274982$-$0506395	&	261.957593	&	$-$5.110977	&	11.076	&	10.281	&	10.081	&	$-$119.02	&	0.22	\\
17275057$-$0504396	&	261.960714	&	$-$5.077668	&	9.395	&	8.384	&	8.089	&	$-$121.31	&	0.31	\\
17275683$-$0504051	&	261.986826	&	$-$5.068087	&	10.676	&	9.788	&	9.546	&	$-$122.43	&	0.18	\\
17275811$-$0501218	&	261.992155	&	$-$5.022726	&	9.949	&	9.006	&	8.718	&	$-$123.42	&	0.23	\\
17280180$-$0507277	&	262.007524	&	$-$5.124366	&	11.130	&	10.329	&	10.111	&	$-$120.88	&	0.23	\\
17280547$-$0502047	&	262.022809	&	$-$5.034661	&	9.596	&	8.572	&	8.309	&	$-$121.48	&	0.28	\\
\hline
\multicolumn{8}{c}{Hydra Probable Non--Members}       \\
\hline
17255789$-$0458546	&	261.491242	&	$-$4.981845	&	10.464	&	9.482	&	9.187	&	$+$40.71	&	0.40	\\
17260595$-$0513249	&	261.524819	&	$-$5.223611	&	11.025	&	10.013	&	9.688	&	$-$117.14	&	0.36	\\
17261100$-$0514439	&	261.545871	&	$-$5.245550	&	10.534	&	9.692	&	9.454	&	$+$43.06	&	0.19	\\
17263463$-$0508564	&	261.644297	&	$-$5.149007	&	11.352	&	10.435	&	10.205	&	$+$25.02	&	0.20	\\
17263699$-$0523021	&	261.654149	&	$-$5.383922	&	10.810	&	9.944	&	9.770	&	$+$54.68	&	0.18	\\
17263709$-$0447537	&	261.654575	&	$-$4.798261	&	10.191	&	9.280	&	8.983	&	$+$23.81	&	0.21	\\
17263955$-$0513427	&	261.664802	&	$-$5.228553	&	10.838	&	9.740	&	9.415	&	$+$105.29	&	0.46	\\
17264946$-$0450003	&	261.706092	&	$-$4.833436	&	10.443	&	9.480	&	9.122	&	$+$67.55	&	0.46	\\
17265327$-$0511019	&	261.721975	&	$-$5.183884	&	11.215	&	10.308	&	10.027	&	$+$96.86	&	0.25	\\
17265679$-$0458163	&	261.736648	&	$-$4.971201	&	9.519	&	8.452	&	8.057	&	$-$78.38	&	2.07	\\
17265821$-$0513464	&	261.742548	&	$-$5.229565	&	10.581	&	9.530	&	9.200	&	$+$12.90	&	0.44	\\
17265901$-$0437299	&	261.745876	&	$-$4.624999	&	9.940	&	8.858	&	8.526	&	$-$104.49	&	0.51	\\
17270635$-$0511447	&	261.776467	&	$-$5.195766	&	10.789	&	9.773	&	9.464	&	$-$42.22	&	0.36	\\
17270794$-$0438448	&	261.783086	&	$-$4.645779	&	10.144	&	9.045	&	8.733	&	$-$148.51	&	0.86	\\
17271236$-$0446567	&	261.801531	&	$-$4.782440	&	11.168	&	10.345	&	10.110	&	$+$58.06	&	4.09	\\
17271261$-$0511574	&	261.802555	&	$-$5.199298	&	11.590	&	10.517	&	10.223	&	$-$46.15	&	0.41	\\
17272208$-$0527013	&	261.842034	&	$-$5.450384	&	10.329	&	9.384	&	9.091	&	$-$92.97	&	0.24	\\
17273428$-$0526069	&	261.892859	&	$-$5.435271	&	11.066	&	10.314	&	10.061	&	$-$78.61	&	0.22	\\
17274666$-$0512061	&	261.944417	&	$-$5.201704	&	9.704	&	8.747	&	8.414	&	$+$71.94	&	0.58	\\
17280295$-$0509455	&	262.012312	&	$-$5.162657	&	11.224	&	10.481	&	10.269	&	$+$55.02	&	0.23	\\
17280791$-$0456141	&	262.032985	&	$-$4.937277	&	10.554	&	9.549	&	9.254	&	$+$53.78	&	0.31	\\
17280997$-$0500041	&	262.041543	&	$-$5.001159	&	11.065	&	10.149	&	9.971	&	$+$9.07	&	0.23	\\
17281257$-$0506430	&	262.052380	&	$-$5.111953	&	9.487	&	8.632	&	8.322	&	$+$7.21	&	0.29	\\
17281658$-$0437371	&	262.069099	&	$-$4.626997	&	10.298	&	9.425	&	9.170	&	$+$30.38	&	0.31	\\
17282141$-$0527043	&	262.089211	&	$-$5.451209	&	10.406	&	9.454	&	9.141	&	$-$39.47	&	0.27	\\
17284000$-$0506557	&	262.166707	&	$-$5.115496	&	10.237	&	9.298	&	8.977	&	$-$30.09	&	0.47	\\
17284713$-$0517356	&	262.196415	&	$-$5.293239	&	10.638	&	9.702	&	9.457	&	$-$44.65	&	0.22	\\
17284827$-$0443511	&	262.201140	&	$-$4.730881	&	11.212	&	10.344	&	10.088	&	$+$65.04	&	0.58	\\
17285076$-$0511360	&	262.211504	&	$-$5.193343	&	10.125	&	9.228	&	8.996	&	$+$31.18	&	0.24	\\
17291922$-$0447570	&	262.330122	&	$-$4.799176	&	10.964	&	10.170	&	9.940	&	$+$15.92	&	0.20	\\
17292260$-$0516203	&	262.344189	&	$-$5.272316	&	11.015	&	10.101	&	9.836	&	$-$15.10	&	0.26	\\
17292502$-$0455071	&	262.354280	&	$-$4.918642	&	10.532	&	9.708	&	9.481	&	$+$10.36	&	0.24	\\
\enddata

\end{deluxetable}

\clearpage
\setlength{\hoffset}{-0.75in}
\tablenum{3}
\tablecolumns{13}
\tablewidth{0pt}

\begin{deluxetable}{ccccccccccccc}
\tabletypesize{\scriptsize}
\tablecaption{NGC 6342 and NGC 6366 Stellar Atmosphere Parameters and 
Abundance Ratios}
\tablehead{
\colhead{Star Name}     &
\colhead{T$_{\rm eff}$}	&
\colhead{log(g)}	&
\colhead{[Fe/H]}	&
\colhead{$\xi$$_{\rm mic.}$}      &
\colhead{[O/Fe]}	&
\colhead{[Na/Fe]}        &
\colhead{[Mg/Fe]}        &
\colhead{[Si/Fe]}        &
\colhead{[Ca/Fe]}        &
\colhead{[Cr/Fe]}        &
\colhead{[Ni/Fe]}        &
\colhead{[La/Fe]}        \\
\colhead{(2MASS)}	&
\colhead{(K)}      &
\colhead{(cgs)}      &
\colhead{(dex)}      &
\colhead{(km s$^{\rm -1}$)}      &
\colhead{(dex)}	&
\colhead{(dex)} &
\colhead{(dex)} &
\colhead{(dex)} &
\colhead{(dex)} &
\colhead{(dex)} &
\colhead{(dex)} &
\colhead{(dex)} 
}

\startdata
\hline
\multicolumn{13}{c}{NGC 6342}       \\
\hline
17205345$-$1957303	&	4600	&	2.20	&	$-$0.41	&	2.05	&	\nodata	&	$-$0.16	&	\nodata	&	$+$0.32	&	$+$0.15	&	\nodata	&	$-$0.04	&	$+$0.28	\\
17210009$-$1935354	&	4200	&	1.50	&	$-$0.59	&	1.70	&	$+$0.45	&	$+$0.17	&	\nodata	&	$+$0.44	&	\nodata	&	\nodata	&	$+$0.00	&	$-$0.05	\\
17210680$-$1935191	&	4850	&	2.75	&	$-$0.49	&	1.55	&	\nodata	&	$-$0.08	&	\nodata	&	$+$0.49	&	$+$0.22	&	\nodata	&	$-$0.04	&	$+$0.31	\\
17211185$-$1934551	&	4175	&	1.40	&	$-$0.65	&	2.20	&	$+$0.77	&	$+$0.02	&	$+$0.37	&	$+$0.41	&	$+$0.29	&	$-$0.13	&	$+$0.10	&	$+$0.12	\\
\hline
\multicolumn{13}{c}{NGC 6366}       \\
\hline
17271061$-$0457415	&	4575	&	2.15	&	$-$0.66	&	2.00	&	$+$0.67	&	$+$0.46	&	$+$0.36	&	$+$0.35	&	$+$0.40	&	$+$0.16	&	$+$0.12	&	$+$0.33	\\
17272071$-$0505087	&	4650	&	2.45	&	$-$0.47	&	1.90	&	$+$0.57	&	$+$0.03	&	$+$0.26	&	$+$0.22	&	$+$0.22	&	$-$0.21	&	$+$0.09	&	$+$0.29	\\
17273010$-$0504197	&	4400	&	1.80	&	$-$0.47	&	1.60	&	$+$0.27	&	$+$0.27	&	$+$0.29	&	$+$0.21	&	$+$0.33	&	$-$0.04	&	\nodata	&	$+$0.14	\\
17273285$-$0500304	&	4375	&	1.75	&	$-$0.63	&	1.70	&	$+$0.60	&	$-$0.11	&	$+$0.30	&	$+$0.35	&	$+$0.17	&	\nodata	&	$+$0.15	&	$+$0.00	\\
17274128$-$0505308	&	4550	&	2.05	&	$-$0.60	&	1.65	&	$+$0.58	&	$+$0.21	&	$+$0.37	&	$+$0.23	&	$+$0.44	&	$-$0.03	&	$+$0.02	&	$+$0.17	\\
17274221$-$0506173	&	4400	&	1.65	&	$-$0.67	&	2.15	&	$+$0.67	&	$+$0.25	&	$+$0.41	&	$+$0.18	&	$+$0.48	&	$+$0.17	&	$+$0.18	&	$+$0.19	\\
17274279$-$0504077	&	4475	&	2.00	&	$-$0.42	&	1.70	&	$+$0.43	&	$+$0.18	&	$+$0.14	&	$+$0.35	&	$+$0.31	&	$-$0.13	&	$+$0.10	&	$+$0.19	\\
17274541$-$0504089	&	4500	&	1.95	&	$-$0.62	&	1.80	&	$+$0.52	&	$+$0.27	&	$+$0.34	&	$+$0.26	&	$+$0.38	&	$+$0.09	&	$+$0.07	&	$+$0.26	\\
17274724$-$0500362	&	4350	&	1.75	&	$-$0.47	&	1.90	&	$+$0.40	&	$+$0.12	&	$+$0.14	&	$+$0.44	&	$+$0.20	&	\nodata	&	$+$0.05	&	$+$0.09	\\
17274809$-$0507395	&	4525	&	2.10	&	$-$0.52	&	1.65	&	$+$0.32	&	$+$0.38	&	$+$0.34	&	$+$0.33	&	$+$0.37	&	$+$0.08	&	$+$0.01	&	$+$0.24	\\
17274982$-$0506395	&	4475	&	1.60	&	$-$0.60	&	1.75	&	$+$0.60	&	$-$0.09	&	$+$0.27	&	$+$0.25	&	$+$0.19	&	$-$0.24	&	$+$0.15	&	$+$0.14	\\
17275683$-$0504051	&	4400	&	1.75	&	$-$0.62	&	1.90	&	$+$0.59	&	$+$0.06	&	$+$0.34	&	$+$0.18	&	$+$0.31	&	$-$0.02	&	$+$0.09	&	$+$0.19	\\
17280180$-$0507277	&	4550	&	2.15	&	$-$0.41	&	1.65	&	$+$0.36	&	$+$0.04	&	$+$0.23	&	$+$0.34	&	$+$0.16	&	\nodata	&	$+$0.14	&	$+$0.13	\\
\enddata

\end{deluxetable}

\clearpage
\setlength{\hoffset}{-0.50in}
\tablenum{4}
\tablecolumns{7}
\tablewidth{0pt}

\begin{deluxetable}{ccccccc}
\tablecaption{Line List and Adopted Reference Abundances}
\tablehead{
\colhead{Species}       &
\colhead{Wavelength}    &
\colhead{E.P.}  &
\colhead{log(gf)\tablenotemark{a}}        &
\colhead{log $\epsilon$(X)$_{\rm \odot}$}       &
\colhead{log $\epsilon$(X)$_{Arc.}$}    &
\colhead{[X/Fe] or [Fe/H]$_{Arc.}$}     \\
\colhead{}      &
\colhead{(\AA)} &
\colhead{(eV)}  &
\colhead{}      &
\colhead{(dex)}      &
\colhead{(dex)}      &
\colhead{(dex)}
}

\startdata
[O I]	&	6300.30	&	0.00	&	$-$9.750	&	8.69	&	8.63	&	$+$0.44	\\
Na I	&	6154.23	&	2.10	&	$-$1.560	&	6.33	&	5.89	&	$+$0.06	\\
Na I	&	6160.75	&	2.10	&	$-$1.210	&	6.33	&	5.89	&	$+$0.06	\\
Mg I	&	6318.71	&	5.10	&	$-$2.010	&	7.58	&	7.38	&	$+$0.30	\\
Mg I	&	6319.24	&	5.10	&	$-$2.250	&	7.58	&	7.38	&	$+$0.30	\\
Mg I	&	6319.49	&	5.10	&	$-$2.730	&	7.58	&	7.38	&	$+$0.30	\\
Si I	&	6142.48	&	5.62	&	$-$1.575	&	7.55	&	7.38	&	$+$0.33	\\
Si I	&	6145.02	&	5.62	&	$-$1.460	&	7.55	&	7.38	&	$+$0.33	\\
Si I	&	6155.13	&	5.62	&	$-$0.774	&	7.55	&	7.38	&	$+$0.33	\\
Si I	&	6155.69	&	5.62	&	$-$2.352	&	7.55	&	7.38	&	$+$0.33	\\
Si I	&	6195.43	&	5.87	&	$-$1.560	&	7.55	&	7.38	&	$+$0.33	\\
Si I	&	6237.32	&	5.61	&	$-$1.115	&	7.55	&	7.38	&	$+$0.33	\\
Si I	&	6244.47	&	5.62	&	$-$1.303	&	7.55	&	7.38	&	$+$0.33	\\
Ca I	&	6122.22	&	1.89	&	$-$0.466	&	6.36	&	6.07	&	$+$0.21	\\
Ca I	&	6156.02	&	2.52	&	$-$2.637	&	6.36	&	6.07	&	$+$0.21	\\
Ca I	&	6161.30	&	2.52	&	$-$1.246	&	6.36	&	6.07	&	$+$0.21	\\
Ca I	&	6162.17	&	1.90	&	$-$0.210	&	6.36	&	6.07	&	$+$0.21	\\
Ca I	&	6166.44	&	2.52	&	$-$1.262	&	6.36	&	6.07	&	$+$0.21	\\
Ca I	&	6169.04	&	2.52	&	$-$0.837	&	6.36	&	6.07	&	$+$0.21	\\
Ca I	&	6169.56	&	2.53	&	$-$0.628	&	6.36	&	6.07	&	$+$0.21	\\
Cr I	&	6330.09	&	0.94	&	$-$3.000	&	5.67	&	5.09	&	$-$0.08	\\
Fe I	&	6094.37	&	4.65	&	$-$1.700	&	7.52	&	7.02	&	$-$0.50	\\
Fe I	&	6100.27	&	4.56	&	$-$2.116	&	7.52	&	7.02	&	$-$0.50	\\
Fe I	&	6151.62	&	2.18	&	$-$3.379	&	7.52	&	7.02	&	$-$0.50	\\
Fe I	&	6159.37	&	4.61	&	$-$1.950	&	7.52	&	7.02	&	$-$0.50	\\
Fe I	&	6165.36	&	4.14	&	$-$1.584	&	7.52	&	7.02	&	$-$0.50	\\
Fe I	&	6173.33	&	2.22	&	$-$2.930	&	7.52	&	7.02	&	$-$0.50	\\
Fe I	&	6180.20	&	2.73	&	$-$2.629	&	7.52	&	7.02	&	$-$0.50	\\
Fe I	&	6187.99	&	3.94	&	$-$1.690	&	7.52	&	7.02	&	$-$0.50	\\
Fe I	&	6200.31	&	2.61	&	$-$2.437	&	7.52	&	7.02	&	$-$0.50	\\
Fe I	&	6219.28	&	2.20	&	$-$2.563	&	7.52	&	7.02	&	$-$0.50	\\
Fe I	&	6229.23	&	2.85	&	$-$2.885	&	7.52	&	7.02	&	$-$0.50	\\
Fe I	&	6232.64	&	3.65	&	$-$1.263	&	7.52	&	7.02	&	$-$0.50	\\
Fe I	&	6240.65	&	2.22	&	$-$3.353	&	7.52	&	7.02	&	$-$0.50	\\
Fe I	&	6252.56	&	2.40	&	$-$1.847	&	7.52	&	7.02	&	$-$0.50	\\
Fe I	&	6253.83	&	4.73	&	$-$1.500	&	7.52	&	7.02	&	$-$0.50	\\
Fe I	&	6270.22	&	2.86	&	$-$2.649	&	7.52	&	7.02	&	$-$0.50	\\
Fe I	&	6315.81	&	4.08	&	$-$1.720	&	7.52	&	7.02	&	$-$0.50	\\
Fe I	&	6322.69	&	2.59	&	$-$2.446	&	7.52	&	7.02	&	$-$0.50	\\
Fe I	&	6330.85	&	4.73	&	$-$1.230	&	7.52	&	7.02	&	$-$0.50	\\
Fe I	&	6335.33	&	2.20	&	$-$2.387	&	7.52	&	7.02	&	$-$0.50	\\
Fe I	&	6336.82	&	3.69	&	$-$0.866	&	7.52	&	7.02	&	$-$0.50	\\
Ni I	&	6128.96	&	1.68	&	$-$3.400	&	6.25	&	5.81	&	$+$0.06	\\
Ni I	&	6130.13	&	4.27	&	$-$1.040	&	6.25	&	5.81	&	$+$0.06	\\
Ni I	&	6175.36	&	4.09	&	$-$0.619	&	6.25	&	5.81	&	$+$0.06	\\
Ni I	&	6176.81	&	4.09	&	$-$0.270	&	6.25	&	5.81	&	$+$0.06	\\
Ni I	&	6177.24	&	1.83	&	$-$3.550	&	6.25	&	5.81	&	$+$0.06	\\
Ni I	&	6186.71	&	4.11	&	$-$0.890	&	6.25	&	5.81	&	$+$0.06	\\
Ni I	&	6191.17	&	1.68	&	$-$2.233	&	6.25	&	5.81	&	$+$0.06	\\
Ni I	&	6223.98	&	4.11	&	$-$0.960	&	6.25	&	5.81	&	$+$0.06	\\
Ni I	&	6322.16	&	4.15	&	$-$1.190	&	6.25	&	5.81	&	$+$0.06	\\
La II	&	6262.29	&	0.40	&	hfs	&	1.13	&	0.57	&	$-$0.06	\\
\enddata

\tablenotetext{a}{The ``hfs" designation indicates the abundance was
calculated by taking hyperfine structure into account.  See text for details.}

\end{deluxetable}

\clearpage
\tablenum{5}
\tablecolumns{10}
\tablewidth{0pt}

\begin{deluxetable}{cccccccccc}
\tabletypesize{\scriptsize}
\tablecaption{NGC 6342 and NGC 6366 Abundance Ratio Uncertainties}
\tablehead{
\colhead{Star Name}     &
\colhead{$\Delta$[Fe/H]}        &
\colhead{$\Delta$[O/Fe]}	&
\colhead{$\Delta$[Na/Fe]}        &
\colhead{$\Delta$[Mg/Fe]}        &
\colhead{$\Delta$[Si/Fe]}        &
\colhead{$\Delta$[Ca/Fe]}        &
\colhead{$\Delta$[Cr/Fe]}        &
\colhead{$\Delta$[Ni/Fe]}        &
\colhead{[$\Delta$La/Fe]}        \\
\colhead{(2MASS)}	&
\colhead{(dex)} &
\colhead{(dex)}	&
\colhead{(dex)} &
\colhead{(dex)} &
\colhead{(dex)} &
\colhead{(dex)} &
\colhead{(dex)} &
\colhead{(dex)} &
\colhead{(dex)} 
}

\startdata
\hline
\multicolumn{10}{c}{NGC 6342}       \\
\hline
17205345$-$1957303	&	0.05	&	\nodata	&	0.09	&	\nodata	&	0.08	&	0.12	&	\nodata	&	0.09	&	0.09	\\
17210009$-$1935354	&	0.06	&	0.10	&	0.10	&	\nodata	&	0.10	&	\nodata	&	\nodata	&	0.13	&	0.10	\\
17210680$-$1935191	&	0.06	&	\nodata	&	0.09	&	\nodata	&	0.09	&	0.13	&	\nodata	&	0.09	&	0.09	\\
17211185$-$1934551	&	0.04	&	0.08	&	0.09	&	0.05	&	0.07	&	0.12	&	0.13	&	0.10	&	0.08	\\
\hline
\multicolumn{10}{c}{NGC 6366}       \\
\hline
17271061$-$0457415	&	0.05	&	0.09	&	0.07	&	0.08	&	0.12	&	0.09	&	0.11	&	0.07	&	0.09	\\
17272071$-$0505087	&	0.05	&	0.09	&	0.06	&	0.07	&	0.08	&	0.09	&	0.11	&	0.07	&	0.09	\\
17273010$-$0504197	&	0.06	&	0.10	&	0.08	&	0.06	&	0.09	&	0.11	&	0.13	&	\nodata	&	0.09	\\
17273285$-$0500304	&	0.06	&	0.10	&	0.07	&	0.06	&	0.09	&	0.11	&	\nodata	&	0.11	&	0.09	\\
17274128$-$0505308	&	0.06	&	0.10	&	0.07	&	0.08	&	0.08	&	0.09	&	0.12	&	0.11	&	0.09	\\
17274221$-$0506173	&	0.05	&	0.09	&	0.07	&	0.08	&	0.08	&	0.10	&	0.12	&	0.08	&	0.09	\\
17274279$-$0504077	&	0.07	&	0.10	&	0.08	&	0.07	&	0.11	&	0.10	&	0.13	&	0.09	&	0.10	\\
17274541$-$0504089	&	0.05	&	0.09	&	0.07	&	0.05	&	0.07	&	0.09	&	0.12	&	0.08	&	0.09	\\
17274724$-$0500362	&	0.06	&	0.10	&	0.08	&	0.08	&	0.09	&	0.11	&	\nodata	&	0.09	&	0.09	\\
17274809$-$0507395	&	0.06	&	0.09	&	0.10	&	0.06	&	0.11	&	0.10	&	0.12	&	0.09	&	0.09	\\
17274982$-$0506395	&	0.07	&	0.10	&	0.10	&	0.07	&	0.08	&	0.10	&	0.12	&	0.10	&	0.10	\\
17275683$-$0504051	&	0.05	&	0.09	&	0.09	&	0.11	&	0.09	&	0.09	&	0.12	&	0.07	&	0.09	\\
17280180$-$0507277	&	0.06	&	0.10	&	0.08	&	0.09	&	0.07	&	0.10	&	\nodata	&	0.09	&	0.10	\\
\enddata

\end{deluxetable}

\clearpage
\tablenum{6}
\tablecolumns{6}
\tablewidth{0pt}

\begin{deluxetable}{lccccc}
\tabletypesize{\scriptsize}
\tablecaption{Composition Comparison: --0.7 $\leq$ [Fe/H] $\leq$ --0.4}
\tablehead{
\colhead{Average [X/Fe]}	&
\colhead{NGC 6342}	&
\colhead{NGC 6366}	&
\colhead{Bulge Clusters\tablenotemark{a}}	&
\colhead{Bulge Clusters\tablenotemark{b}}      &
\colhead{Bulge Field}      
}

\startdata
$<$[O/Fe]$>$	&	 $+$0.61	&	 $+$0.51	&	 $+$0.48	&	 $+$0.16	&	 $+$0.47	\\
$<$[Na/Fe]$>$	&	 $-$0.01	&	 $+$0.16	&	 $+$0.17	&	 $+$0.36	&	 $-$0.01	\\
$<$[Mg/Fe]$>$	&	 $+$0.37	&	 $+$0.29	&	 $+$0.39	&	 $+$0.34	&	 $+$0.31	\\
$<$[Si/Fe]$>$	&	 $+$0.42	&	 $+$0.28	&	 $+$0.29	&	 $+$0.31	&	 $+$0.29	\\
$<$[Ca/Fe]$>$	&	 $+$0.22	&	 $+$0.30	&	 $+$0.31	&	 $+$0.22	&	 $+$0.26	\\
$<$[Cr/Fe]$>$	&	 $-$0.13	&	 $-$0.02	&	 $-$0.07	&	 $-$0.08	&	 $+$0.01	\\
$<$[Ni/Fe]$>$	&	 $+$0.01	&	 $+$0.10	&	 $-$0.04	&	 $+$0.02	&	 $+$0.06	\\
$<$[La/Fe]$>$	&	 $+$0.17	&	 $+$0.18	&	 \nodata	&	 $+$0.36	&	 $-$0.20	\\
\hline
\multicolumn{6}{c}{Abundance Dispersions}       \\
\hline
$\sigma$[O/Fe]	&	0.23	&	0.13	&	0.07	&	0.32	&	0.15	\\
$\sigma$[Na/Fe]	&	0.14	&	0.17	&	0.15	&	0.25	&	0.12	\\
$\sigma$[Mg/Fe]	&	\nodata	&	0.08	&	0.07	&	0.11	&	0.09	\\
$\sigma$[Si/Fe]	&	0.07	&	0.08	&	0.09	&	0.12	&	0.11	\\
$\sigma$[Ca/Fe]	&	0.07	&	0.11	&	0.10	&	0.16	&	0.11	\\
$\sigma$[Cr/Fe]	&	\nodata	&	0.14	&	0.03	&	0.18	&	0.10	\\
$\sigma$[Ni/Fe]	&	0.07	&	0.05	&	0.05	&	0.09	&	0.05	\\
$\sigma$[La/Fe]	&	0.17	&	0.09	&	\nodata	&	0.09	&	0.08	\\
\enddata

\tablenotetext{a}{NGC 6388 and NGC 6441 are omitted.}
\tablenotetext{b}{All bulge globular clusters with --0.7 $\leq$ [Fe/H] $\leq$
--0.4 are included.}

\end{deluxetable}

\clearpage
\tablenum{7}
\tablecolumns{2}
\tablewidth{0pt}

\begin{deluxetable}{ll}
\tabletypesize{\scriptsize}
\tablecaption{Literature References}
\tablehead{
\colhead{Stellar Population}     &
\colhead{Reference}
}

\startdata
HP--1	&	Barbuy et al. (2006)	\\
NGC 6342	&	Origlia et al. (2005a)	\\
NGC 6352	&	Feltzing et al. (2009)	\\
NGC 6388	&	Carretta et al. (2007)	\\
NGC 6388	&	Carretta et al. (2009a)	\\
NGC 6388	&	Worley \& Cottrell (2010)	\\
NGC 6440	&	Origlia et al. (2008)	\\
NGC 6441	&	Gratton et al. (2006)	\\
NGC 6441	&	Gratton et al. (2007)	\\
NGC 6441        &       Origlia et al. (2008)   \\
NGC 6528	&	Carretta et al. (2001)	\\
NGC 6528	&	Origlia et al. (2005a)	\\
NGC 6539	&	Origlia et al. (2005b)	\\
NGC 6553	&	Cohen et al. (1999)	\\
NGC 6553	&	Mel{\'e}ndez et al. (2003)	\\
NGC 6553	&	Alves$-$Brito et al. (2006)	\\
NGC 6553	&	Johnson et al. (2014)	\\
NGC 6569	&	Valenti et al. (2011)	\\
NGC 6624	&	Valenti et al. (2011)	\\
Terzan 5	&	Origlia et al. (2011)	\\
Terzan 5	&	Origlia et al. (2013)	\\
UKS 1	&	Origlia et al. (2005b)	\\
Galactic Bulge	&	Alves$-$Brito et al. (2010)	\\
Galactic Bulge	&	Gonzalez et al. (2011)	\\
Galactic Bulge	&	Hill et al. (2011)	\\
Galactic Bulge	&	Johnson et al. (2011)	\\
Galactic Bulge	&	Johnson et al. (2012)	\\
Galactic Bulge	&	Bensby et al. (2013)	\\
Galactic Bulge	&	Johnson et al. (2013)	\\
Galactic Bulge	&	Johnson et al. (2014)	\\
\enddata

\end{deluxetable}

%\clearpage
%\setlength{\voffset}{0.90in}
%\input{tab4.tex}
%\clearpage
%\setlength{\hoffset}{-0.75in}
%\input{tab5.tex}
%\clearpage
%\setlength{\voffset}{0.90in}
%\input{tab3.tex}
%\clearpage
%\setlength{\voffset}{0.90in}
%\input{tab4.tex}
%\clearpage
%\setlength{\voffset}{0.90in}
%\input{tab5.tex}


\begin{thebibliography}{}

\bibitem[Alonso et al.(1997)]{1997A&A...323..374A} Alonso, A., Salaris, M., Martinez-Roger, C., Straniero, O., \& Arribas, S.\ 1997, \aap, 323, 374

\bibitem[Alonso-Garc{\'{\i}}a et al.(2012)]{2012AJ....143...70A} 
Alonso-Garc{\'{\i}}a, J., Mateo, M., Sen, B., et al.\ 2012, \aj, 143, 70

\bibitem[Alves-Brito et al.(2006)]{2006A&A...460..269A} Alves-Brito, A., Barbuy, B., Zoccali, M., et al.\ 2006, \aap, 460, 269

\bibitem[Alves-Brito et al.(2010)]{2010A&A...513A..35A} Alves-Brito, A., Mel{\'e}ndez, J., Asplund, M., Ram{\'{\i}}rez, I., \& Yong, D.\ 2010, \aap, 513, AA35

\bibitem[Arnould et al.(1999)]{1999A&A...347..572A} Arnould, M., Goriely, S., \& Jorissen, A.\ 1999, \aap, 347, 572

\bibitem[Barbuy et al.(2006)]{2006A&A...449..349B} Barbuy, B., Zoccali, M., Ortolani, S., et al.\ 2006, \aap, 449, 349

\bibitem[Barbuy et al.(2009)]{2009A&A...507..405B} Barbuy, B., Zoccali, M., Ortolani, S., et al.\ 2009, \aap, 507, 405

\bibitem[Bellini et al.(2013)]{2013ApJ...765...32B} Bellini, A., Piotto, 
G., Milone, A.~P., et al.\ 2013, \apj, 765, 32

\bibitem[Bensby et al.(2011)]{2011A&A...533A.134B} Bensby, T., Ad{\'e}n, D., Mel{\'e}ndez, J., et al.\ 2011, \aap, 533, A134

\bibitem[Bensby et al.(2013)]{2013A&A...549A.147B} Bensby, T., Yee, J.~C., Feltzing, S., et al.\ 2013, \aap, 549, A147

\bibitem[Bershady et al.(2008)]{2008SPIE.7014E..0HB} Bershady, M., Barden, 
S., Blanche, P.-A., et al.\ 2008, \procspie, 7014, 70140H

\bibitem[Bica et al.(2006)]{2006A&A...450..105B} Bica, E., Bonatto, C., Barbuy, B., \& Ortolani, S.\ 2006, \aap, 450, 105

\bibitem[Busso et al.(1999)]{1999ARA&A..37..239B} Busso, M., Gallino, R., \& Wasserburg, G.~J.\ 1999, \araa, 37, 239

\bibitem[Campos et al.(2013)]{2013MNRAS.433..243C} Campos, F., Kepler, 
S.~O., Bonatto, C., \& Ducati, J.~R.\ 2013, \mnras, 433, 243

\bibitem[Carretta et al.(2001)]{2001AJ....122.1469C} Carretta, E., Cohen, 
J.~G., Gratton, R.~G., \& Behr, B.~B.\ 2001, \aj, 122, 1469

\bibitem[Carretta et al.(2004)]{2004A&A...416..925C} Carretta, E., Gratton, R.~G., Bragaglia, A., Bonifacio, P., \& Pasquini, L.\ 2004, \aap, 416, 925

\bibitem[Carretta et al.(2007)]{2007A&A...464..967C} Carretta, E., Bragaglia, A., Gratton, R.~G., et al.\ 2007, \aap, 464, 967

\bibitem[Carretta et al.(2009a)]{2009A&A...505..117C} Carretta, E., Bragaglia, A., Gratton, R.~G., et al.\ 2009a, \aap, 505, 117

\bibitem[Carretta et al.(2009b)]{2009A&A...505..139C} Carretta, E., Bragaglia, A., Gratton, R., \& Lucatello, S.\ 2009b, \aap, 505, 139

\bibitem[Carretta et al.(2014)]{2014A&A...564A..60C} Carretta, E., Bragaglia, A., Gratton, R.~G., et al.\ 2014, \aap, 564, A60

\bibitem[Carretta(2015)]{2015ApJ...810..148C} Carretta, E.\ 2015, \apj, 
810, 148 

\bibitem[Casetti-Dinescu et al.(2010)]{2010AJ....140.1282C} 
Casetti-Dinescu, D.~I., Girard, T.~M., Korchagin, V.~I., van Altena, W.~F., 
\& L{\'o}pez, C.~E.\ 2010, \aj, 140, 1282

\bibitem[Castelli \& Kurucz(2004)]{2004astro.ph..5087C} Castelli, F., \& Kurucz, R.~L.\ 2004, arXiv:astro-ph/0405087

\bibitem[Cohen et al.(1999)]{1999ApJ...523..739C} Cohen, J.~G., Gratton, 
R.~G., Behr, B.~B., \& Carretta, E.\ 1999, \apj, 523, 739

\bibitem[Cohen \& Kirby(2012)]{2012ApJ...760...86C} Cohen, J.~G., \& Kirby, E.~N.\ 2012, \apj, 760, 86

\bibitem[Cordero et al.(2014)]{2014ApJ...780...94C} Cordero, M.~J.,
Pilachowski, C.~A., Johnson, C.~I., et al.\ 2014, \apj, 780, 94

\bibitem[Cordero et al.(2015)]{2015ApJ...800....3C} Cordero, M.~J.,
Pilachowski, C.~A., Johnson, C.~I., \& Vesperini, E.\ 2015, \apj, 800, 3

\bibitem[C{\^o}t{\'e}(1999)]{1999AJ....118..406C} C{\^o}t{\'e}, P.\ 1999, 
\aj, 118, 406

\bibitem[Da Costa \& Seitzer(1989)]{1989AJ.....97..405D} Da Costa, G.~S., \& Seitzer, P.\ 1989, \aj, 97, 405

\bibitem[Da Costa \& Armandroff(1995)]{1995AJ....109.2533D} Da Costa, G.~S., \& Armandroff, T.~E.\ 1995, \aj, 109, 2533

\bibitem[D'Orazi et al.(2010)]{2010ApJ...719L.213D} D'Orazi, V., Gratton, R., Lucatello, S., et al.\ 2010, \apjl, 719, L213

\bibitem[Dotter et al.(2008)]{2008ApJS..178...89D} Dotter, A., Chaboyer, 
B., Jevremovi{\'c}, D., et al.\ 2008, \apjs, 178, 89

\bibitem[Dotter et al.(2010)]{2010ApJ...708..698D} Dotter, A., Sarajedini, 
A., Anderson, J., et al.\ 2010, \apj, 708, 698

\bibitem[Dubath et al.(1997)]{1997A&A...324..505D} Dubath, P., Meylan, G., \& Mayor, M.\ 1997, \aap, 324, 505

\bibitem[Feltzing et al.(2009)]{2009A&A...493..913F} Feltzing, S., Primas, F., \& Johnson, R.~A.\ 2009, \aap, 493, 913

\bibitem[Ferraro et al.(2009)]{2009Natur.462..483F} Ferraro, F.~R., 
Dalessandro, E., Mucciarelli, A., et al.\ 2009, \nat, 462, 483

\bibitem[Forbes \& Bridges(2010)]{2010MNRAS.404.1203F} Forbes, D.~A., \& Bridges, T.\ 2010, \mnras, 404, 1203

\bibitem[Freeman \& Norris(1981)]{1981ARA&A..19..319F} Freeman, K.~C., \& Norris, J.\ 1981, \araa, 19, 319

\bibitem[Gonz{\'a}lez Hern{\'a}ndez \& Bonifacio(2009)]{2009A&A...497..497G} Gonz{\'a}lez Hern{\'a}ndez, J.~I., \& Bonifacio, P.\ 2009, \aap, 497, 497

\bibitem[Gonzalez et al.(2011)]{2011A&A...530A..54G} Gonzalez, O.~A., Rejkuba, M., Zoccali, M., et al.\ 2011, \aap, 530, A54

\bibitem[Gratton et al.(2004)]{2004ARA&A..42..385G} Gratton, R., Sneden, C., \& Carretta, E.\ 2004, \araa, 42, 385

\bibitem[Gratton et al.(2012)]{2012A&ARv..20...50G} Gratton, R.~G., Carretta, E., \& Bragaglia, A.\ 2012, \aapr, 20, 50

\bibitem[Gratton et al.(2006)]{2006A&A...455..271G} Gratton, R.~G., Lucatello, S., Bragaglia, A., et al.\ 2006, \aap, 455, 271

\bibitem[Gratton et al.(2007)]{2007A&A...464..953G} Gratton, R.~G., Lucatello, S., Bragaglia, A., et al.\ 2007, \aap, 464, 953

\bibitem[Harris(1996)]{1996AJ....112.1487H} Harris, W.~E.\ 1996, \aj, 112,
1487

\bibitem[Heitsch \& Richtler(1999)]{1999A&A...347..455H} Heitsch, F., \& Richtler, T.\ 1999, \aap, 347, 455

\bibitem[Hill et al.(2011)]{2011A&A...534A..80H} Hill, V., Lecureur, A., G{\'o}mez, A., et al.\ 2011, \aap, 534, A80

\bibitem[Hinkle et al.(2000)]{2000vnia.book.....H} Hinkle, K., Wallace, L.,
Valenti, J., \& Harmer, D.\ 2000, Visible and Near Infrared Atlas of the Arcturus Spectrum 3727-9300 A ed.~Kenneth Hinkle, Lloyd Wallace, Jeff Valenti, and Dianne Harmer.~(San Francisco: ASP) ISBN: 1-58381-037-4, 2000

\bibitem[James et al.(2004)]{2004A&A...427..825J} James, G., Fran{\c c}ois, P., Bonifacio, P., et al.\ 2004, \aap, 427, 825

\bibitem[Johnson et al.(1982)]{1982ApJ...258..161J} Johnson, H.~R., Mould, 
J.~R., \& Bernat, A.~P.\ 1982, \apj, 258, 161

\bibitem[Johnson \& Pilachowski(2010)]{2010ApJ...722.1373J} Johnson, C.~I., \& Pilachowski, C.~A.\ 2010, \apj, 722, 1373

\bibitem[Johnson et al.(2011)]{2011ApJ...732..108J} Johnson, C.~I., Rich,
R.~M., Fulbright, J.~P., Valenti, E., \& McWilliam, A.\ 2011, \apj, 732, 108

\bibitem[Johnson et al.(2012)]{2012ApJ...749..175J} Johnson, C.~I., Rich,
R.~M., Kobayashi, C., \& Fulbright, J.~P.\ 2012, \apj, 749, 175

\bibitem[Johnson et al.(2013)]{2013ApJ...765..157J} Johnson, C.~I., Rich,
R.~M., Kobayashi, C., et al.\ 2013, \apj, 765, 157

\bibitem[Johnson et al.(2014)]{2014AJ....148...67J} Johnson, C.~I., Rich, 
R.~M., Kobayashi, C., Kunder, A., \& Koch, A.\ 2014, \aj, 148, 67

\bibitem[Johnson et al.(2015)]{2015AJ....150...63J} Johnson, C.~I., Rich, 
R.~M., Pilachowski, C.~A., et al.\ 2015, \aj, 150, 63

\bibitem[Knezek et al.(2010)]{2010SPIE.7735E..7DK} Knezek, P.~M., Bershady, 
M.~A., Willmarth, D., et al.\ 2010, \procspie, 7735, 77357D

\bibitem[Kraft(1994)]{1994PASP..106..553K} Kraft, R.~P.\ 1994, \pasp, 106, 
553

\bibitem[Kunder et al.(2012)]{2012AJ....143...57K} Kunder, A., Koch, A., 
Rich, R.~M., et al.\ 2012, \aj, 143, 57

\bibitem[Langer et al.(1993)]{1993PASP..105..301L} Langer, G.~E., Hoffman, R., \& Sneden, C.\ 1993, \pasp, 105, 301

\bibitem[Lawler et al.(2001)]{2001ApJ...556..452L} Lawler, J.~E., 
Bonvallet, G., \& Sneden, C.\ 2001, \apj, 556, 452

\bibitem[Mar{\'{\i}}n-Franch et al.(2009)]{2009ApJ...694.1498M} 
Mar{\'{\i}}n-Franch, A., Aparicio, A., Piotto, G., et al.\ 2009, \apj, 694, 
1498

\bibitem[Mauro et al.(2012)]{2012ApJ...761L..29M} Mauro, F., Moni Bidin, 
C., Cohen, R., et al.\ 2012, \apjl, 761, L29

\bibitem[McWilliam \& Rich(1994)]{1994ApJS...91..749M} McWilliam, A., \& Rich, R.~M.\ 1994, \apjs, 91, 749

\bibitem[McWilliam et al.(2010)]{2010IAUS..265..279M} McWilliam, A., 
Fulbright, J., \& Rich, R.~M.\ 2010, IAU Symposium, 265, 279

\bibitem[Mel{\'e}ndez et al.(2003)]{2003A&A...411..417M} Mel{\'e}ndez, J., Barbuy, B., Bica, E., et al.\ 2003, \aap, 411, 417

\bibitem[Mel{\'e}ndez et al.(2008)]{2008A&A...484L..21M} Mel{\'e}ndez, J., Asplund, M., Alves-Brito, A., et al.\ 2008, \aap, 484, L21

\bibitem[Minniti(1995)]{1995AJ....109.1663M} Minniti, D.\ 1995, \aj, 109, 
1663

\bibitem[Nataf et al.(2013)]{2013ApJ...766...77N} Nataf, D.~M., Gould, 
A.~P., Pinsonneault, M.~H., \& Udalski, A.\ 2013, \apj, 766, 77

\bibitem[Ness et al.(2013)]{2013MNRAS.432.2092N} Ness, M., Freeman, K., 
Athanassoula, E., et al.\ 2013, \mnras, 432, 2092

\bibitem[O'connell et al.(2011)]{2011PASP..123.1139O} O''connell, J.~E.,
Johnson, C.~I., Pilachowski, C.~A., \& Burks, G.\ 2011, \pasp, 123, 1139

\bibitem[Origlia et al.(2005a)]{2005MNRAS.356.1276O} Origlia, L., Valenti, 
E., \& Rich, R.~M.\ 2005a, \mnras, 356, 1276

\bibitem[Origlia et al.(2005b)]{2005MNRAS.363..897O} Origlia, L., Valenti, 
E., Rich, R.~M., \& Ferraro, F.~R.\ 2005b, \mnras, 363, 897

\bibitem[Origlia et al.(2008)]{2008MNRAS.388.1419O} Origlia, L., Valenti, 
E., \& Rich, R.~M.\ 2008, \mnras, 388, 1419

\bibitem[Origlia et al.(2011)]{2011ApJ...726L..20O} Origlia, L., Rich, 
R.~M., Ferraro, F.~R., et al.\ 2011, \apjl, 726, L20

\bibitem[Origlia et al.(2013)]{2013ApJ...779L...5O} Origlia, L., Massari, 
D., Rich, R.~M., et al.\ 2013, \apjl, 779, L5

\bibitem[Ortolani et al.(1997)]{1997MNRAS.284..692O} Ortolani, S., Bica, 
E., \& Barbuy, B.\ 1997, \mnras, 284, 692

\bibitem[Ortolani et al.(2011)]{2011ApJ...737...31O} Ortolani, S., Barbuy, 
B., Momany, Y., et al.\ 2011, \apj, 737, 31

\bibitem[Paust et al.(2009)]{2009AJ....137..246P} Paust, N.~E.~Q., 
Aparicio, A., Piotto, G., et al.\ 2009, \aj, 137, 246

\bibitem[Piotto et al.(2015)]{2015AJ....149...91P} Piotto, G., Milone, 
A.~P., Bedin, L.~R., et al.\ 2015, \aj, 149, 91

\bibitem[Prantzos et al.(2007)]{2007A&A...470..179P} Prantzos, N., Charbonnel, C., \& Iliadis, C.\ 2007, \aap, 470, 179

\bibitem[Ram{\'{\i}}rez \& Allende Prieto(2011)]{2011ApJ...743..135R} Ram{\'{\i}}rez, I., \& Allende Prieto, C.\ 2011, \apj, 743, 135

\bibitem[Rich et al.(1997)]{1997ApJ...484L..25R} Rich, R.~M., Sosin, C., 
Djorgovski, S.~G., et al.\ 1997, \apjl, 484, L25

\bibitem[Roederer \& Thompson(2015)]{2015MNRAS.449.3889R} Roederer, I.~U., \& Thompson, I.~B.\ 2015, \mnras, 449, 3889

\bibitem[Rossi et al.(2015)]{2015MNRAS.450.3270R} Rossi, L.~J., Ortolani, 
S., Barbuy, B., Bica, E., \& Bonfanti, A.\ 2015, \mnras, 450, 3270

\bibitem[Rutledge et al.(1997)]{1997PASP..109..883R} Rutledge, G.~A., 
Hesser, J.~E., Stetson, P.~B., et al.\ 1997, \pasp, 109, 883

\bibitem[Ryde et al.(2010)]{2010A&A...509A..20R} Ryde, N., Gustafsson, B., Edvardsson, B., et al.\ 2010, \aap, 509, A20

\bibitem[Sarajedini et al.(2007)]{2007AJ....133.1658S} Sarajedini, A., 
Bedin, L.~R., Chaboyer, B., et al.\ 2007, \aj, 133, 1658

\bibitem[Saviane et al.(2012)]{2012A&A...540A..27S} Saviane, I., da Costa, G.~S., Held, E.~V., et al.\ 2012, \aap, 540, A27

\bibitem[Skrutskie et al.(2006)]{2006AJ....131.1163S} Skrutskie, M.~F.,
Cutri, R.~M., Stiening, R., et al.\ 2006, \aj, 131, 1163

\bibitem[Sneden(1973)]{1973ApJ...184..839S} Sneden, C.\ 1973, \apj, 184, 839

\bibitem[Sneden et al.(2008)]{2008ARA&A..46..241S} Sneden, C., Cowan, J.~J., \& Gallino, R.\ 2008, \araa, 46, 241

\bibitem[Sneden et al.(2014)]{2014ApJS..214...26S} Sneden, C., Lucatello, 
S., Ram, R.~S., Brooke, J.~S.~A., \& Bernath, P.\ 2014, \apjs, 214, 26

\bibitem[Szentgyorgyi et al.(2011)]{2011PASP..123.1188S} Szentgyorgyi, A., 
Furesz, G., Cheimets, P., et al.\ 2011, \pasp, 123, 1188

\bibitem[Tinsley(1979)]{1979ApJ...229.1046T} Tinsley, B.~M.\ 1979, \apj, 
229, 1046

\bibitem[Valenti et al.(2004)]{2004MNRAS.351.1204V} Valenti, E., Ferraro, 
F.~R., \& Origlia, L.\ 2004, \mnras, 351, 1204

\bibitem[Valenti et al.(2011)]{2011MNRAS.414.2690V} Valenti, E., Origlia, 
L., \& Rich, R.~M.\ 2011, \mnras, 414, 2690

\bibitem[VandenBerg et al.(2013)]{2013ApJ...775..134V} VandenBerg, D.~A., 
Brogaard, K., Leaman, R., \& Casagrande, L.\ 2013, \apj, 775, 134

\bibitem[van den Bergh(2003)]{2003ApJ...590..797V} van den Bergh, S.\ 2003, 
\apj, 590, 797

\bibitem[Ventura et al.(2012)]{2012ApJ...761L..30V} Ventura, P., D'Antona, F., Di Criscienzo, M., et al.\ 2012, \apjl, 761, L30

\bibitem[Worley \& Cottrell(2010)]{2010MNRAS.406.2504W} Worley, C.~C., \& Cottrell, P.~L.\ 2010, \mnras, 406, 2504

\bibitem[Yong et al.(2005)]{2005A&A...438..875Y} Yong, D., Grundahl, F., Nissen, P.~E., Jensen, H.~R., \& Lambert, D.~L.\ 2005, \aap, 438, 875

\bibitem[Yong et al.(2014)]{2014MNRAS.441.3396Y} Yong, D., Roederer, I.~U., 
Grundahl, F., et al.\ 2014, \mnras, 441, 3396

\bibitem[Zinn(1985)]{1985ApJ...293..424Z} Zinn, R.\ 1985, \apj, 293, 424

\bibitem[Zoccali et al.(2014)]{2014A&A...562A..66Z} Zoccali, M., Gonzalez, O.~A., Vasquez, S., et al.\ 2014, \aap, 562, A66

\end{thebibliography}
\end{document}